\documentclass[10pt]{article}
\usepackage{graphs}
\usepackage[round]{natbib}
\usepackage{amsmath}
\usepackage{amsthm}
\usepackage{amsfonts}
\usepackage{amssymb}
\usepackage{mathrsfs}
\usepackage{textcomp}
\usepackage{epsf}
\usepackage[dvips]{graphicx}
\usepackage[latin1]{inputenc}
\usepackage{xspace}
\usepackage[T1]{fontenc}
\usepackage{epsfig,color}
\usepackage{booktabs}

\author{Frédéric Fourré$^{1}$\thanks{Present Address: Systems Biology Group, Life Sciences Research Unit, University of Luxembourg, 162a, avenue de la Fa\"{\i}encerie, L-1511 Luxembourg, Grand-Duchy of Luxembourg. Correspondence should be addressed to frederic.fourre@uni.lu} \and Denis Baurain$^{2}$\thanks{Present Address: Unit of Animal Genomics, GIGA-R and Faculty of Veterinary Medicine, University of Liège, B34, B-4000 Liège, Belgium}} 

\title{A reduction method for noisy Boolean networks} 

\newtheoremstyle{exmp}
  {3pt}
  {3pt}
  {\small \upshape}
  {}
  {\normalsize \bfseries}
  {}
  {0.5em}
  {}

\theoremstyle{exmp}
\newtheorem{exmp}{Example}

\newtheoremstyle{rem}
  {3pt}
  {3pt}
  {\small \itshape}
  {}
  {\normalsize \bfseries}
  {}
  {0.5em}
  {}

\theoremstyle{rem}
\newtheorem{rem}{Remark}

\newtheoremstyle{prop}
  {3pt}
  {3pt}
  {\normalsize \itshape}
  {}
  {\normalsize \bfseries}
  {}
  {0.5em}
  {}

\theoremstyle{prop}
\newtheorem{prop}{Proposition}

\begin{document}


\maketitle 

\noindent
$^{1}$Affiliation: Montefiore Institute, Department of Applied Mathematics, University of Liège, B28, B-4000 Liège, Belgium \\
$^{2}$Affiliation: Algology, Mycology and Experimental Systematics, Department
of Life Sciences, University of Liège, B22, B-4000 Liège, Belgium

\begin{abstract}
This paper is concerned with the reduction of a noisy synchronous Boolean network to a coarse-grained Markov chain model. Consider an $n$-node Boolean network having at least two basins of attraction and where each node may be perturbed with probability $0<p<1$ during one time step. This process is a discrete-time homogeneous Markov chain with $2^{n}$ possible states. Now under certain conditions, the transitions between the basins of the network may be approximated by a homogeneous Markov chain where each state of the chain represents a basin of the network, i.e. the size of the reduced chain is the number of attractors of the original network. 
\end{abstract}

\tableofcontents 
 
\section{Noisy Boolean networks (\textsf{NBN}s)} 

\subsection{Definition}

Boolean networks (\textsf{BNs}) have been used for several decades as models of biochemical networks, mainly to predict their qualitative properties and, in the case of genetic regulatory networks, to infer the inputs and interaction rules of their nodes from microarray data \citep{KA69,GK73,KA93,LI04,MA07,SJ08}.

An $n$-node \textsf{BN} model consists of $n$ interacting nodes which, in the context of biochemical networks, generally represent genes or molecular species such as proteins, RNAs or metabolites. Let $X_{i}(k)$ denote the Boolean variable describing the state of node $i$ at discrete time $k=0,1,\ldots$ If node $i$ represents a protein, then we say that if the value of the node is $1$, then the protein is present, in its active form and its target (or substrate) present. Using De Morgan's law, the negation of this conjunction gives the interpretation for value $0$: the protein is absent or inactive or its target (or substrate) absent. 

\begin{rem}
A more concise interpretation instead of ``present'' (resp. ``absent'') is ``present and not being degraded'' (resp. ``absent or being degraded'') or even ``present and production rate greater than degradation rate'' (resp. ``absent or degradation rate greater than production rate''). 
\end{rem}

For an $n$-node synchronous \textsf{BN}, the interactions between the nodes are modeled by a set of $n$ Boolean interaction functions such that:

\begin{equation}\label{eqintfct}
X_{i}(k+1)=F_{i}[X_{1}(k),X_{2}(k),\ldots,X_{n}(k)], \quad i=1,2,\ldots,n,
\end{equation} with $F_{i}$ interaction function of node $i$. If the state of node $i$ at time $(k+1)$ depends on the state of node $j$ at previous time $k$, then node $j$ is said to be an input for node $i$. The number of inputs of a node is called the connectivity of that node. The network is said to be synchronous because as expressed in (\ref{eqintfct}) the $n$ nodes are updated synchronously. Also from (\ref{eqintfct}) the dynamics of the network deterministic.

Let us take an example to illustrate some essential properties of \textsf{BN}s. Consider the following \textsf{BN}:

\begin{eqnarray}\label{eqbooex}
X_{1}(k+1) & = & \neg{X}_{4}(k) \nonumber \\
X_{2}(k+1) & = & \neg{X}_{3}(k) \nonumber \\
X_{3}(k+1) & = & X_{1}(k) \vee X_{2}(k) \vee X_{4}(k) \nonumber \\
X_{4}(k+1) & = & X_{1}(k) \wedge X_{2}(k) \wedge X_{3}(k),
\end{eqnarray} where symbols $\neg$, $\vee$ and $\wedge$ represent logical operators NOT, OR and AND respectively. The size of the state space of this network, i.e. the number of possible states, is $2^{n}=16$ since $n=4$ here. From (\ref{eqbooex}), the next state for each possible state can be computed to obtain the state table of the network. Using this state table then, the state diagram can be built. This is shown in Fig \ref{figbooex} where it can be seen that the state space has been partioned into two disjoint sets. These sets are called basins of attraction and are denoted by \textsf{B1} and \textsf{B2} in the figure. Each basin consists of an attractor and some transient states. The attractor of \textsf{B1} is the fixed point $1010$: whatever the initial state in \textsf{B1}, the network will converge to $1010$ and stay there forever. Attractors may also be periodic as is the case for basin \textsf{B2}. The size of a basin or of an attractor is the number of states that constitute it. 

\begin{figure}
\begin{center}
\begin{graph}(12,6)(-6,-3)

\textnode{b1}(-4,0){\textsf{0110}}[\graphlinecolour{1}]
\textnode{b2}(-2.3,0){\textsf{1010}}[\graphlinecolour{1}]
\textnode{b3}(-3.2,1){\textsf{0101}}[\graphlinecolour{1}]
\textnode{b4}(-4.8,1){\textsf{1101}}[\graphlinecolour{1}]
\textnode{b5}(-4.8,-1){\textsf{0001}}[\graphlinecolour{1}]
\textnode{b6}(-3.2,-1){\textsf{1001}}[\graphlinecolour{1}]
\diredge{b1}{b2}[\grapharrowlength{0.18}\grapharrowwidth{0.8}]
\diredge{b3}{b1}[\grapharrowlength{0.18}\grapharrowwidth{0.8}]
\diredge{b4}{b1}[\grapharrowlength{0.18}\grapharrowwidth{0.8}]
\diredge{b5}{b1}[\grapharrowlength{0.18}\grapharrowwidth{0.8}]
\diredge{b6}{b1}[\grapharrowlength{0.18}\grapharrowwidth{0.8}]
\dirloopedge{b2}(0.5,0)(0.4,0.5)[\grapharrowlength{0.18}\grapharrowwidth{0.8}]
\textnode{b7}(2.3,0){\textsf{1000}}[\graphlinecolour{1}]
\textnode{b8}(4.3,0){\textsf{1011}}[\graphlinecolour{1}]
\textnode{b9}(3.3,1){\textsf{1110}}[\graphlinecolour{1}]
\textnode{b10}(3.3,-1){\textsf{0010}}[\graphlinecolour{1}]
\dirbow{b7}{b9}{0.2071}[\grapharrowlength{0.18}\grapharrowwidth{0.8}]
\dirbow{b9}{b8}{0.2071}[\grapharrowlength{0.18}\grapharrowwidth{0.8}]
\dirbow{b8}{b10}{0.2071}[\grapharrowlength{0.18}\grapharrowwidth{0.8}]
\dirbow{b10}{b7}{0.2071}[\grapharrowlength{0.18}\grapharrowwidth{0.8}]
\textnode{b11}(5,1){\textsf{0100}}[\graphlinecolour{1}]
\textnode{b12}(1.6,1){\textsf{1100}}[\graphlinecolour{1}]
\textnode{b13}(-0.1,1){\textsf{0000}}[\graphlinecolour{1}]
\textnode{b15}(1.6,-1){\textsf{0011}}[\graphlinecolour{1}]
\textnode{b14}(5,-1){\textsf{0111}}[\graphlinecolour{1}]
\textnode{b16}(-0.1,-1){\textsf{1111}}[\graphlinecolour{1}]
\diredge{b11}{b9}[\grapharrowlength{0.18}\grapharrowwidth{0.8}]
\diredge{b12}{b9}[\grapharrowlength{0.18}\grapharrowwidth{0.8}]
\diredge{b13}{b12}[\grapharrowlength{0.18}\grapharrowwidth{0.8}]
\diredge{b14}{b10}[\grapharrowlength{0.18}\grapharrowwidth{0.8}]
\diredge{b15}{b10}[\grapharrowlength{0.18}\grapharrowwidth{0.8}]
\diredge{b16}{b15}[\grapharrowlength{0.18}\grapharrowwidth{0.8}]
\freetext(-4,-2){\textsf{B1}}
\freetext(3.3,-2){\textsf{B2}}
\end{graph}
\end{center}
\caption{\textsf{State diagram of the four-node network defined by interaction rules (\ref{eqbooex}). Interactions between the nodes lead to the partitioning of the state space into two basins B1 and B2 of respective sizes $6$ and $10$}.}
\label{figbooex} 
\end{figure}

Now suppose that, due to random perturbations, each node of a \textsf{BN} has probability $0 < p < 1$ to switch its state ($0$ to $1$ or vice versa) between any two times $k$ and $(k+1)$. Then we get a noisy \textsf{BN} (\textsf{NBN})\footnote{Here, for the sake of simplicity, perturbation probability $p$ is supposed to be independent of time $k$ and of the states of the nodes. If, for instance, for a particular node, $0$ to $1$ random switching probability is set greater than $1$ to $0$ one, then it means that random perturbations tend to turn the node ON rather than OFF.}. The introduction of disorder $p$ is supported by the stochastic nature of intracellular processes coupled to the fact that, thermodynamically speaking, biochemical networks are open systems. 

One important difference between a \textsf{BN} and a \textsf{NBN} is that in the latter, transitions between the basins of the network are allowed. For example, in the state diagram of Fig.~\ref{figbooex}, if we perturb simultaneously nodes $2$ and $3$ of attractor state $1010$ which is in \textsf{B1}, then we go to transient state $1100$ which belongs to \textsf{B2}. 

\begin{rem}
Flipping the value of single nodes has been envisaged by \citet{KA93} in genetic Boolean networks to study their stability to what he called minimal perturbations. \citet{SH02A} proposed a model for random gene perturbations based on probabilistic Boolean networks (a class of Boolean networks that enclosed the class of synchronous Boolean networks by assigning more than one interaction function to each node) in which any gene of the network may flip with probability $p$ independently of other genes and interpreted the perturbation events as the influence of external stimuli on the activity of the genes.
\end{rem}

Let $L(x;n,p)$ be the probability that exactly $x$ nodes will be perturbed during one time step. Then letting $q=1-p$:

\begin{equation}\label{eqlxmp}
L(x;n,p)= \left\{ \begin{array}{ll} 
\binom{n}{x} p^{x}q^{n-x} & \textrm{if} \quad 0<x<n, \\
q^{n} &  \textrm{if} \quad x=0, \\
p^{n} &  \textrm{if} \quad x=n.
\end{array} \right.
\end{equation} This is the binomial probability distribution with parameters $n$ and $p$. On average, $np$ nodes are perturbed at each time step. The probability that at least one node will be perturbed during one time step is therefore:

\begin{equation}\label{eqqm}
\sum_{x=1}^{n} L(x;n,p)=1-q^{n}=r.
\end{equation} Since we are mainly interested in the behaviour of the network as $p$ varies, in the following $n$ and $x$ will be considered as parameters while $p$ will be considered as a variable. Thus we should write $L(p;n,x)$ instead of $L(x;n,p)$\footnote{$L(p;n,x)$ is a function of $p$ with parameters $n$ and $x$. It is not a probability density function. For fixed $p$, $L(p;n,x)$ is a probability.}.

For $0<x<n$, $L(p;n,x)$ has a maximum at $p=x/n$ and $L(0;n,x)=L(1;n,x)=0$. 

The Maclaurin series of $L(p;n,x)$ up to order $2$ is: 

\[
L(p;n,x)= \left\{ \begin{array}{ll} 
1 - np + n(n-1) p^{2}/2 + \ldots & \textrm{if} \quad x=0, \\
np - n(n-1) p^{2} + \ldots & \textrm{if} \quad x=1, \\
n(n-1) p^{2}/2 + \ldots &  \textrm{if} \quad x=2, \\
0+\ldots &  \textrm{if} \quad 2 < x \leq n.
\end{array} \right.
\] Thus one has:

\begin{equation}\label{eqapproxL}
L(p;n,x)= \left\{ \begin{array}{ll} 
1-np + o(p) & \textrm{if} \quad x=0, \\
np + o(p) &  \textrm{if} \quad x=1, \\
o(p) &  \textrm{if} \quad 2 \leq x \leq n.
\end{array} \right.
\end{equation} This means that, $n$ being fixed, for sufficiently small $p$ the probability that $2 \leq x \leq n$ nodes be perturbed during one time step is negligible compared to the probability that just one node be perturbed. Table~\ref{tablpmx} gives some values of $L(p;n,x)$ rounded to four decimal places. We see that at $p=0.002$, $L(p;4,1)=0.0080=4p$ and $L(p;8,1)=0.0158 \approx 8p$. 

\begin{table}
\begin{center}
\begin{tabular}{@{}ccccc@{}}
\toprule
        & \multicolumn{4}{c}{$p$} \\ \cmidrule(l){2-5}
        & 0.002  & 0.02 & 0.2 & 0.5  \\ \midrule
$L(p;4,1)$ & 0.0080 &    0.0753 &     0.4096 &    0.2500 \\
$L(p;4,2)$ & 0.0000 &    0.0023 &     0.1536 &    0.3750 \\
$L(p;4,3)$ & 0.0000 &    0.0000 &     0.0256 &    0.2500 \\
$L(p;4,4)$ & 0.0000 &    0.0000 &     0.0016 &    0.0625 \\
$L(p;8,1)$ & 0.0158 &    0.1389 &     0.3355 &    0.0312 \\
$L(p;8,2)$ & 0.0001 &    0.0099 &     0.2936 &    0.1094 \\
$L(p;8,3)$ & 0.0000 &    0.0004 &     0.1468 &    0.2188 \\
$L(p;8,4)$ & 0.0000 &    0.0000 &     0.0459 &    0.2734 \\
\bottomrule
\end{tabular} 
\caption{\textsf{Some values of the function $L(p;n,x)$}.}
\label{tablpmx}
\end{center}
\end{table}

\subsection{The mean specific path}

Consider a series of Bernoulli trials where each time step defines one trial and where a success means that at least one node has been perturbed. From (\ref{eqqm}), the probability that the first success will occur on trial $i$ is:

\begin{equation}\label{eqgeo} 
p_{i} = r (1-r)^{i-1}, \quad i=1,2,\ldots,
\end{equation} which is a geometric distribution with parameter $r$. The first moment of this distribution (the mean time between two successes) is:

\begin{equation}\label{eqmoygeo}
\tau=\frac{1}{r}=\frac{1}{1-q^{n}}.
\end{equation} We see that $\tau$ is a decreasing function of $p$ and $n$, that tends to $1$ as $p \to 1$ and to $\infty$ as $p \to 0$. Since time step is unity, $\tau$ equals the mean number $d$ of state transitions between two successes. By analogy with the mean free path of a particle in physics\footnote{In kinetic theory of gases, the mean free path of a gas molecule is the mean distance traveled by the molecule between two successive collisions.}, $d$ will be called \textbf{mean specific path}. The term ``specific'' is used to recall that between two successes, the trajectory in the state space is specific to node interactions, i.e., it is entirely determined by node interactions.

For sufficiently small $p$ one has $r=1-q^{n} \approx np \approx L(p;n,1)$, where the last approximation comes from (\ref{eqapproxL}). Hence, for sufficiently small $p$ we may write:

\[
p_{i} \approx np (1-np)^{i-1}, \quad i=1,2,\ldots
\] 

Depending on the value of $p$, different dynamical regimes are possible. For $p=0$, the network is trapped by an attractor where it stays forever. For $0 < p < 1$, transitions between the basins of the network occur. The more $p$ is close to one, the shorter the times spent on the attractors, the more the network suffers from functional instability. In the low $p$ regime, the network may both maintain a specific activity for a long time period and change its activity: functional stability and flexibility (or diversity) coexist.

\subsection{Time evolution equation}

A \textsf{NBN} is in fact a discrete-time Markov chain $\{ \mathbf{X}_{k}, k=0,1,\ldots \}$, where $\mathbf{X}_{k}$ is the random variable representing the state of the network at time $k$. The state space of an $n$-node \textsf{BN} will be denoted by $\{ 1,2,\ldots,E \}$, with $E=2^{n}$ and with state $i$ corresponding to binary representation of $(i-1)$.

Let $\pi_{ij}=\mathsf{Pr} \{ \mathbf{X}_{k+1}=j \vert \mathbf{X}_{k}=i \} \geq 0$ be the conditional probability that the network will be in state $j$ at $(k+1)$ while in state $i$ at $k$. The matrix of size $E$ whose elements are the $\pi_{ij}$'s will be denoted by $\Pi$ and is called the transition probability matrix in Markov theory. The sum of elements in each row of $\Pi$ is unity. Let $z_{i}^{(k)}$ be the probability that the network will be in state $i$ at time $k$ and denote by $\mathbf{z}^{(k)}$ the vector whose elements are the state probabilities $z_{i}^{(k)}$. Given an initial state probability vector $\mathbf{z}^{(0)}$, the vectors $\mathbf{z}^{(1)},\mathbf{z}^{(2)},\ldots$ are found from \citep{KL75}:

\begin{equation}\label{eqmarkov} 
\mathbf{z}^{(k+1)} = \mathbf{z}^{(k)} \Pi, \quad k=0,1,\ldots
\end{equation} 

Matrix $\Pi$ is the sum of two matrices:

\begin{enumerate}
\item The perturbation matrix $\Pi'$, whose $(i,j)$th element is equal to

\[
\pi_{ij}' = \left\{ \begin{array}{ll} 
p^{h_{ij}}q^{n-h_{ij}} & \textrm{if} \quad i \neq j, \\
0 &  \textrm{if} \quad i=j,
\end{array} \right.
\] where $h_{ij}$ refers to $(i,j)$th element of Hamming distance matrix $H$, a symmetric matrix that depends only on $n$. Element $h_{ij}$ is equal to the number of bits that differ in the Boolean representations of states $i$ and $j$. Diagonal elements of $H$ are therefore $0$. For example, if $n=2$, $H$ takes the form:

\begin{equation}
H =
\left( \begin{array}{cccccccc}
     0 &    1 &    1 &    2 \\
     1 &    0 &    2 &    1 \\
     1 &    2 &    0 &    1 \\
     2 &    1 &    1 &    0 \\
\end{array} \right).
\end{equation} Note that in each row of $H$, integers $x=1,2,\ldots,n$ appear respectively $\binom{n}{x}$ times. Thus from (\ref{eqlxmp}) and (\ref{eqqm}), the sum of elements in each row of $\Pi'$ must be $r$. 


\item The interaction matrix $\Pi''$. If the node interactions are such that starting in state $i$ the network is forced to transition to state $j$ in one time step, then $(i,j)$th element of $\Pi''$ equals $q^{n}=1-r$, otherwise it is $0$. Notice that if $i$th diagonal element of $\Pi''$ equals $q^{n}$ then state $i$ is a fixed point (attractor of size $1$).
\end{enumerate}

Therefore, for fixed $n$, any probability $\pi_{ij}$ will be either $0$ or a function of $p$.

\subsection{Stationary probability distribution}

Since $p$ does not depend on $k$, the $\pi_{ij}$'s are independent of time. The chain is thus homogeneous. As shown by \citet{SH02A} for genetic probabilistic Boolean networks, for $0<p<1$, the chain is irreducible and aperiodic. Consequently, for given $p$ there is a unique stationary distribution $\bar{\mathbf{z}}$ which is independent of $\mathbf{z}^{(0)}$ \citep{KL75}. The stationary distribution satisfies the two equations:

\begin{equation}\label{eqpi}
\mathbf{z} = \mathbf{z} \Pi \quad \textrm{and} \quad \sum_{i} z_{i}=1.
\end{equation} In addition, we have:

\begin{equation}\label{eqpistat}
\lim_{k \to \infty} \Pi^{k} = \bar{\Pi},
\end{equation} where each row of $\bar{\Pi}$ is equal to vector $\bar{\mathbf{z}}$.

\begin{figure}
\begin{center}
\includegraphics[width=1\textwidth]{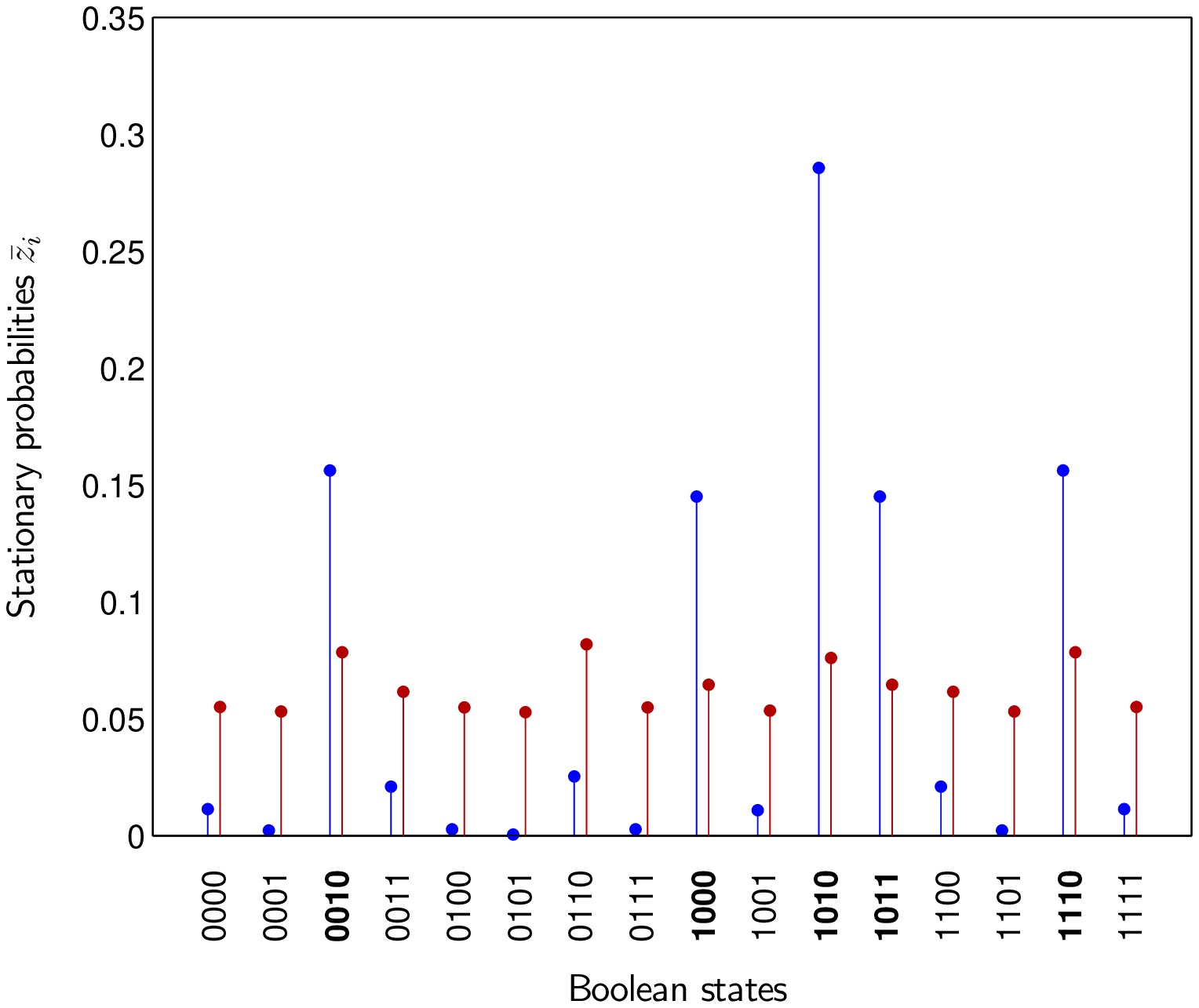}
\caption{\textsf{Stationary state probabilities for the network shown in Fig.~\ref{figbooex} and two $p$ values. Blue points: $p=0.04$; red points: $p=0.4$. Attractor states are in bold type. The stationary probability of B1 (obtained by summing the stationary probabilities of the states of B1) is $0.3270$ when $p=0.04$ and $0.3708$ when $p=0.4$. Since the size of B1 is $6$, this probability converges to $6/16=0.375$ as $p$ tends to $1$}.}
\label{figthermo}
\end{center}
\end{figure}

Fig. \ref{figthermo} shows the stationary state probabilities $\bar{z}_{i}$ for the network of Fig. \ref{figbooex} and two $p$ values. Blue points correspond to $p=0.04$ and red ones to $p=0.4$. As can be seen in the figure, when $p=0.04$,  the stationary probabilities of the transient states are small compared to those of the attractor states (attractor states are represented in bold type in the figure). Therefore in the low $p$ regime, the activity of the network in the long term is governed mostly by its attractors\footnote{Equivalently, in the low $p$ regime, it is mostly attractors that determine the fate of the cells.}. In fact, the stationary probabilities of the transient states tend to $0$ as $p$ tends to $0$. Also notice for $p=0.04$ the stationary probabilities of the second attractor almost equal. 

As $p$ tends to unity, the stationary probabilities tend to be uniform, i.e. as $p \to 1$ one has $\bar{z}_{i} \to 1/E=1/16=0.0625$ $\forall i$ (see the case $p=0.4$). Thus for $p$ sufficiently close to unity, the stationary probabilities of the transient states are not negligible compared to those of the attractor states. 

\begin{rem}
For sufficiently small $p$, the stationary probability of an attractor state may be approximated by $\bar{Z}^{*}/A$ where $\bar{Z}^{*}$ is the limit as $p \to 0$ of the stationary occupation probability $\bar{Z}$ of the basin containing the attractor and $A$ the size of the attractor (see further 4.3). Thus in the case of the network of Fig. \ref{figbooex}, we get for attractor state $11$ (fixed point) that $\bar{z}_{11} \approx 1/3=\lim_{p \to 0} \bar{Z}/1$, and for the states of the second attractor, that each stationary state probability $\approx 1/6=\lim_{p \to 0} \bar{Z}/4$. 
\end{rem}

\section{Method for the calculation of the sojourn time distribution in a basin of a \textsf{NBN}}

Consider a \textsf{BN} having at least two basins. Consider a basin \textsf{B} of size $B$ of this network and a fixed perturbation probability $p$. Let

\begin{equation}\label{eqremabs}
\Pi_{b}=
\left( \begin{array}{cc}
Q & \mathbf{a} \\
\mathbf{0} & 1 \\
\end{array} \right).
\end{equation} The square $B \times B$ sub-matrix $Q$ contains the one-step transition probabilities $\pi_{ij}$ between the states of \textsf{B}. For example, in the case of basin \textsf{B1} of Fig.~\ref{figbooex}, element $(1,4)$ of $Q$ is equal to the one-step transition probability between state $2$ ($0001$) and state $10$ ($1001$). Element $i$ of vector $\mathbf{a}$ represents the one-step transition probability between state $i \in \textsf{B}$ and an absorbing state\footnote{A state such that once reached, it is not possible to escape from it.} regrouping states $j \notin \textsf{B}$, i.e.

\[
a_{i}=\sum_{j \notin \textsf{B}} \pi_{ij}= \sum_{j \notin \textsf{B}} \pi_{ij}'=\sum_{j \notin \textsf{B}} p^{h_{ij}} q^{n-h_{ij}}.
\] $\mathbf{0}$ is a row vector of length $B$ with all elements zero and $1$ is a scalar. 

From matrix $Q$, we can calculate the probability $W^{(k)}$ that the network will be in basin \textsf{B} at time $k$ given an initial probability vector $\mathbf{b}^{(0)}$ of length $B$ with sum of elements $1$:

\begin{equation}\label{eqmarkovb} 
\mathbf{b}^{(k+1)} = \mathbf{b}^{(k)} Q, \quad k=0,1,\ldots, 
\end{equation} and by definition:

\[
 W^{(k)}=\sum_{i=1}^{B} b^{(k)}_{i}, \quad k=0,1,\ldots,
\] with $W^{(0)}=1$. 

Now we want the sojourn time distribution in \textsf{B}. We shall denote by $S$ the discrete random variable representing the sojourn time in \textsf{B}, with sample space $s=1,2,\ldots$, and by $\psi_{s}$ the probability distribution of $S$, i.e. $\psi_{s}=\mathsf{Pr} \{ S=s \}$. For a given basin, $\psi_{s}$ depends on $p$ and initial vector $\mathbf{b}^{(0)}$. It is given by:

\begin{equation}\label{eqpsi} 
\psi_{s} = W^{(s-1)} - W^{(s)}, \quad s=1,2,\ldots,
\end{equation} while the cumulative distribution function of $S$ is:

\begin{equation}\label{eqcumpsi} 
\hat{\psi}_{s} = \mathsf{Pr} \{ S \leq s \} = 1 - W^{(s)}, \quad s=1,2,\ldots
\end{equation}


Matrix $\Pi_{b}$ in (\ref{eqremabs}) represents an absorbing Markov chain \citep{SN59}, i.e. it has at least one absorbing state and from every non-absorbing state (every state $\in \textsf{B}$) one can reach an absorbing state (any state $\notin \textsf{B}$). If the chain is initially in state $i \in \textsf{B}$, then the mean time spent in state $j \in \textsf{B}$ before absorption is the $(i,j)$th element of fundamental matrix $(I-Q)^{-1}$ \citep{SN59}. Thus if $\mu$ denotes the mean of $\psi_{s}$ then:

\begin{equation}\label{eqmuq}
\mu= \mathbf{b}^{(0)} (I-Q)^{-1} \mathbf{1}.
\end{equation} Here $\mathbf{1}$ is a column vector of length $B$ with each element $1$.

Also notice from the definitions of the mean of $S$ and of $\psi_{s}$ that $\mu$ can be expressed as follows:

\[
\mu= \sum_{k=0}^{\infty} W^{(k)}=1+W^{(1)}+W^{(2)}+\ldots
\]

\section{The problem of geometric approximation}

Let $g_{s}$ be a geometric distribution with parameter $p_{\mu}=1/ \mu$, i.e. $\psi_{s}$ and $g_{s}$ have same mean. Consider the maximum deviation between the cumulative distribution functions (cdfs) of these two distributions, namely:

\begin{equation}\label{eqdelmax} 
\delta^{*} = \max_{s \geq 1} \delta^{(s)},
\end{equation} with $\delta^{(s)}=\vert \hat{\psi}_{s} - \hat{g}_{s} \vert$ and $\hat{g}_{s}$ the cdf of $g_{s}$, i.e. $\hat{g}_{s}=1-(1-p_{\mu})^{s}$, $s=1,2,\ldots$ In probability theory, $\delta^{*}$ is called the Kolmogorov metric. As $\psi_{s}$, $\delta^{*}$ is not defined for $p=0$ and, for a given basin, depends on $p$ and $\mathbf{b}^{(0)}$. It comes from (\ref{eqcumpsi}) that:

\begin{equation}\label{eqdels} 
\delta^{(s)} = \vert (1-p_{\mu})^{s} - W^{(s)} \vert, \quad s=1,2,\ldots
\end{equation} Thus $\delta^{(s)}$ represents the absolute error between $W^{(s)}$ and its geometric approximation $(1-p_{\mu})^{s}$. For basins of size $B=1$ (one fixed point and no transient states), $\psi_{s}$ is given by (\ref{eqgeo}) which is geometric. Hence for such basins $\delta^{(s)}=0$ $\forall s,p$ and thus $\delta^{*}=0$ $\forall p$. This is a particular case and in the following we shall study the behaviour of $\delta^{*}$ as $p$ tends to $0$ without any assumptions on $\psi_{s}$.

The probability to exit \textsf{B} during $(k,k+1)$ is:

\begin{equation}\label{eqps12}
p_{e}(k,k+1)= \mathsf{Pr} \{ S=k+1 \vert S > k \}=1- \frac{W^{(k+1)}}{W^{(k)}}, \quad k=0,1,\ldots
\end{equation} This probability depends on $k$, $p$ and $\mathbf{b}^{(0)}$. In particular, for $B=1$, $p_{e}(k,k+1)$ is constant and equal to $r$. When $p >0$, any state $i \in \textsf{B}$ communicates with any state $j \notin \textsf{B}$, thus $p_{e}(k,k+1) >0$ $\forall k=0,1,\ldots$ and therefore probability $W^{(k)}$ is strictly decreasing.

Inversely, since $W^{(0)}=1$, we have:

\begin{equation}\label{eqwn}
W^{(s)} = \prod_{k=0}^{s-1} [1- p_{e}(k,k+1)], \quad s=1,2,\ldots
\end{equation} and thus 

\begin{equation}\label{eqpsihat}
\delta^{(s)} = \vert (1-p_{\mu})^{s} - \prod_{k=0}^{s-1} [1- p_{e}(k,k+1)] \vert, \quad s=1,2,\ldots
\end{equation} We see that if $p_{e}(k,k+1)=p_{\mu}$ $\forall k$ then $\delta^{*}=0$. Also note the following:

\begin{enumerate}
\item
\begin{equation}\label{eqlimmu}
\lim_{p \to 0} \mu = \infty \quad \forall \mathbf{b}^{(0)}.
\end{equation} Whatever the initial conditions, as $p$ becomes smaller, it takes on average more and more time to leave the basin. Also this means that $\forall \mathbf{b}^{(0)}$: $p_{\mu} \to 0$ as $p \to 0$.

\item

\begin{equation}\label{eqlimpe}
\lim_{p \to 0} p_{e}(k,k+1) = 0 \quad \forall k, \mathbf{b}^{(0)}. 
\end{equation} Whatever $k$ and the initial conditions, the probability to leave the basin during $(k,k+1)$ goes to $0$ as $p \to 0$. Indeed, from equations (\ref{eqmarkovb}) and (\ref{eqps12}), it comes that

\begin{equation}\label{eqps13}
p_{e}(k,k+1) = \sum_{i \in \textsf{B}} a_{i} \hat{b}_{i}^{(k)} = \sum_{i \in \textsf{B}} \sum_{x=1}^{n} \Gamma_{i}^{x} p^{x}q^{n-x} \hat{b}_{i}^{(k)}, \quad k=0,1,\ldots,
\end{equation} with $\Gamma_{i}^{x} \geq 0$ the number of ways of leaving \textsf{B} by perturbing $x$ bits of state $i \in \textsf{B}$\footnote{This number is equal to the number of elements in row $i$ of the Hamming distance matrix $H$ that are equal to $x$ and whose column index $j$ is such that $j \notin \textsf{B}$.} and

\[
\hat{b}_{i}^{(k)}=\frac{b_{i}^{(k)}}{W^{(k)}},
\] i.e.

\[
\sum_{i \in \textsf{B}} \hat{b}_{i}^{(k)}=1, \quad k=0,1,\ldots
\] 
\end{enumerate}

In order to show that $\forall \mathbf{b}^{(0)}$ any $\delta^{(s)}$ tends to $0$ as $p$ tends to $0$, we introduce two propositions.

\begin{prop}
For any basin of a noisy Boolean network with fixed $0 < p < 1$, the fundamental matrix $(I-Q)^{-1}$ has a real simple eigenvalue $\lambda^{*} > 1$ which is greater in modulus than any other eigenvalue modulus, that is $\lambda^{*} > \vert \lambda \vert$ for any other eigenvalue $\lambda$ of $(I-Q)^{-1}$. We have:

\[
\lim_{k \to \infty} p_{e}(k,k+1)=\frac{1}{\lambda^{*}}.
\] For a given basin, the rate of convergence depends on $p$ and $\mathbf{b}^{(0)}$ while the limit $1/\lambda^{*}$ depends only on $p$.
\end{prop}

From that proposition, we see that if $\psi_{s}$ is geometric then $p_{\mu}=1/\lambda^{*}$. For example, if $B=1$ then $p_{\mu}=r=1/\lambda^{*}$.

Demonstration.

We have to show that:

\[
\lim_{k \to \infty} \frac{W^{(k+1)}}{W^{(k)}} = \lambda_{b},
\] with $\lambda_{b}=1-1/\lambda^{*}$ a real simple eigenvalue of $Q$ greater in modulus than any other eigenvalue modulus and $0 < \lambda_{b} <1$. 

Matrix $Q$ is nonnegative. It is irreducible and aperiodic thus primitive. From the Perron-Frobenius theorem, we have that: (1) $Q$ has a real eigenvalue $\lambda_{b} > 0$ which is greater in modulus than any other eigenvalue modulus. Since $Q$ is substochastic, we have $0 < \lambda_{b} < 1$. (2) $\lambda_{b}$ is a simple root of the characteristic equation of $Q$. Moreover \citep{DB99}:

\[
\lim_{k \to \infty} \frac{Q^{k}}{\lambda_{b}^{k}}= v u^{T},
\] with $v$ and $u$ right and left eigenvectors associated with $\lambda_{b}$ chosen in such a way that $u>0$, $v>0$ and $u^{T} v=1$. The greater $k$, the better the approximation $Q^{k} \approx \lambda_{b}^{k} v u^{T}$. Thus:

\[
W^{(k+1)} = \mathbf{b}^{(0)} Q^{k+1} \mathbf{1} \approx \mathbf{b}^{(0)} \lambda_{b}^{k+1} v u^{T} \mathbf{1} \approx  \mathbf{b}^{(0)} \lambda_{b} Q^{k} \mathbf{1} = \lambda_{b} W^{(k)}.
\] Therefore:

\[
\frac{W^{(k+1)}}{W^{(k)}} \to \lambda_{b} \quad \textrm{as} \quad k \to \infty.
\]

\begin{prop}
$\mu$ is asymptotically equivalent to $\lambda^{*}$:

\[
\lim_{p \to 0} \frac{\mu}{\lambda^{*}}= 1.
\] 
\end{prop}

Since $\lambda^{*}$ does not depend on $\mathbf{b}^{(0)}$, the last statement is equivalent to say that as $p$ tends to $0$, $\mu$ is less and less dependent on $\mathbf{b}^{(0)}$. In other words, from (\ref{eqmuq}), the elements of vector $(I-Q)^{-1} \mathbf{1}$ tends to be equal as $p \to 0$. 

Demonstration.

If $\psi_{s}$ is geometric then $\mu$ does not depend on $\mathbf{b}^{(0)}$. Thus from (\ref{eqmuq}) it comes that:

\[
(I-Q)^{-1} \mathbf{1} = \mu \mathbf{1},
\] or equivalently

\begin{equation}\label{eqq1}
Q \mathbf{1} = (1-\frac{1}{\mu}) \mathbf{1}.
\end{equation} This means that $(1-1 /\mu)$ is an eigenvalue of $Q$ with associated eigenvector $\mathbf{1}$. Since $Q$ is nonnegative primitive, by the Perron-Frobenius theorem, $(1-\frac{1}{\mu})$ must equal $\lambda_{b}$, which implies $\mu >1$. Now $\lambda_{b}=1-1/ \lambda^{*}$. Thus $\mu=\lambda^{*}$. From (\ref{eqq1}) we see that the sum of elements in any row of $Q$ is a constant. This is because when $\psi_{s}$ is geometric, the probability to leave the basin from any state is constant. 
 

When $Q$ is not geometric, the Perron-Frobenius theorem gives:

\[
Q v = \lambda_{b} v, \quad 0 < \lambda_{b} < 1.
\] Now as $p$ tends to $0$, $Q$ tends to a stochastic matrix (because vector $\mathbf{a}$ tends to vector null). Therefore as $p$ tends to $0$, $\lambda_{b}$ must tend to $1$ and $v$ must tend to $\mathbf{1}$. Thus for sufficiently small $p$ we may write:

\begin{equation}\label{eqqlamb}
Q \mathbf{1} \approx \lambda_{b} \mathbf{1},
\end{equation} or equivalently

\[
(I-Q)^{-1} \mathbf{1} \approx \lambda^{*} \mathbf{1}.
\] Therefore

\[
\mu = \mathbf{b}^{(0)} (I-Q)^{-1} \mathbf{1} \approx \mathbf{b}^{(0)} \lambda^{*} \mathbf{1} = \lambda^{*}.
\] Thus as $p \to 0$, $\mu$ will be less and less dependent on initial conditions and the error in the above approximation will tend to $0$. Note that since $\lambda_{b} \to 1$ as $p \to 0$ we must have $\mu \to \infty$ as $p \to 0$ which is result (\ref{eqlimmu}). 

\begin{rem}
As will be discussed later, approximation $\mu \approx \lambda^{*}$ may be good in some neighborhood of some $p$ (typically in the neighborhood of $p=0.5$, see Fig.~13 in 4.3). 
\end{rem}

From Proposition 2 now, $(1-p_{\mu})^s$ will tend to $(1-1/ \mu)^s=\lambda_{b}^s$ as $p \to 0$. On the other hand, from (\ref{eqqlamb}), we get

\[
Q^{s} \mathbf{1} \approx \lambda_{b}^{s} \mathbf{1},
\] and thus:

\[
W^{(s)}=\mathbf{b}^{(s)} \mathbf{1} = \mathbf{b}^{(0)} Q^{s} \mathbf{1} \approx \lambda_{b}^{s}.
\] Hence from (\ref{eqdels}) any $\delta^{(s)}$ will tend to $0$ as $p$ tends to $0$, whatever $\mathbf{b}^{(0)}$. Thus the maximum will do so:

\[
\lim_{p \to 0} \delta^{*} = 0 \quad \forall \mathbf{b}^{(0)}.
\] Since the maximum deviation between the two cdfs of $\psi_{s}$ and $g_{s}$ vanishes as $p \to 0$, these two cdfs tend to coincide as p tends to $0$. Also this means that $(1-p_{\mu})^{s}$ is a good approximation to within $\pm \delta^{*}$ of $W^{(s)}$ at any and every $s$ and that the error can be made arbitrarily small by taking $p$ sufficiently close to $0$ (but not equal to $0$). 

In addition, from (\ref{eqlimmu}) we may write for sufficiently small $p$:

\[
\hat{g}_{s} = \mathsf{Pr} \{ S \leq s \} = 1-e^{s \ln (1-p_{\mu})} \approx 1-e^{- p_{\mu} s}.
\] Yet the cdf of an exponential distribution with parameter $\gamma$ is:

\[
\mathsf{Pr} \{ T \leq t \} =1 - e^{- \gamma t}.
\] For $t=s$ and $\gamma=p_{\mu}$, the two probabilities are equal. Therefore, as $p \to 0$ and $\forall \mathbf{b}^{(0)}$, the cdf of $\psi_{s}$ tends to coincide with the cdf of an exponential distribution. Since $\mu$ is less and less dependent on $\mathbf{b}^{(0)}$ as $p \to 0$ (see Proposition~2), one has for sufficiently small $p$ that an exponential distribution can be used to approximate the time spent in a basin of a \textsf{NBN}.
 
\begin{rem}
Depending on $\mathbf{b}^{(0)}$, $\psi_{s}$ may be very close to a geometric distribution while $p$ is not small. See Examples~1 and 2 below.
\end{rem}

Suppose that for each basin of a \textsf{NBN} the sojourn time $S$ is sufficiently memoryless. Then, under some additional assumptions that will be discussed in section 5, one may define another discrete-time homogeneous Markov chain $\tilde{\mathbf{Y}}_{k}$ whose states are the basins of the network. (1) This new stochastic process is coarser than $\mathbf{X}_{k}$ since only transitions between the basins of the network are described. (2) The size of its state space is in general much smaller than $E$, the size of the state space of the original process. (3) It has to be viewed as an approximation. The tilde symbol in $\tilde{\mathbf{Y}}_{k}$ is used to recall that in general, a basin does not retain the Markov property (i.e., $\psi_{s}$ is not geometric).

\begin{exmp}
\emph{Calculation of sojourn time distributions and illustration of the geometric approximation problem}. Let us take basin \textsf{B1} of Fig.~\ref{figbooex} with $p=0.02$ and $b_{i}^{(0)}=1/6$ $\forall i=1,2,\ldots,6$, and let us compute $\psi_{s}$ and $g_{s}$. The two distributions are plotted in Fig.~\ref{figmar1}a. The circles stand for the probabilities $\psi_{s}$ ($\mu=13.3530$), the points for the probabilities $g_{s}$ ($p_{\mu}=1 / \mu=0.0749$). Fig. \ref{figmar1}b shows $W^{(k)}$ and its geometric approximation $(1-p_{\mu})^{k}$ versus time $k$. Maximum deviation $\delta^{*}$ between the two is $2.3007\%$.  

Table \ref{tabdis} gives $\mu$ and $\delta^{*}$ for different $p$ values and different initial conditions for the two basins of Fig. \ref{figbooex}. Each of the four columns below $\mu$ and $\delta^{*}$ corresponds to one $p$ value, from left to right: $p=0.002$, $0.02$, $0.2$ and $0.8$. The mean values of the columns are also given (see the rows with ``Mean''). For the calculation of $\mu$, we considered two types of initial conditions: (1) the network starts in one state of the basin (successively each state of the basin was taken as initial state) and (2) $b_{i}^{(0)}=1/B$ $\forall i=1,2,\ldots,B$ (see ``Unif.'' in the Table). We see that for the first three values of $p$ the mean sojour times in \textsf{B1} are smaller than those in \textsf{B2} (see the first three columns below $\mu$). Thus \textsf{B1} is less stable than \textsf{B2}. Also calculated for each basin and the four $p$ values the variation coefficient of the sojourn times obtained with the first type of initial conditions. The results are (columns $1$, $2$, $3$ and $4$): $0.1876$, $1.7880$, $9.3454$ and $15.9039 \%$ for \textsf{B1}; $0.1566$, $1.3030$, $2.5152$ and $17.0044\%$ for \textsf{B2}. Therefore $\mu$ is less and less sensitive to the initial state as $p$ decreases, which is in accordance with Proposition~2.

Another important observation in this table is that, depending on the initial conditions, $\psi_{s}$ may be close to a geometric distribution while $p$ is not small (see in the table the values of $\delta^{*}$ when $p=0.2$). Fig. \ref{figdelp} shows the functions $\delta^{*}(p)$ for the two basins of Fig. \ref{figbooex} and the uniform initial condition. We see that in both cases $\delta^{*}(p) \to 0$ when $p \to 0$, $\delta^{*}(0.5)=0$ ($\psi_{s}$ is geometric at $p=0.5$)\footnote{Whatever the initial conditions, at $p=0.5$, $\mu=8/5$ for \textsf{B1} and $8/3$ for \textsf{B2}. See 5.1.2 for analytical expression of $\mu$ when $p=0.5$.} and $\delta^{*}(p)$ tends to a global maximum as $p \to 1$. For basin \textsf{B1} (blue curve), there is one local maximum at $p=0.112$, while for basin \textsf{B2} (green curve) there are two local maxima, one at $p=0.0530$ and the other at $p=0.3410$, and one local minimum at $p=0.2600$ ($0.0677\%$). As mentioned above, function $\delta^{*}(p)$ depends on initial conditions. For example, if the network starts in state $0000$, then \textsf{B2} has still one local minimum but this now occurs at $p=0.3640$ ($0.0830\%$). 

\begin{figure}
\begin{center}
\includegraphics[width=0.8\textwidth]{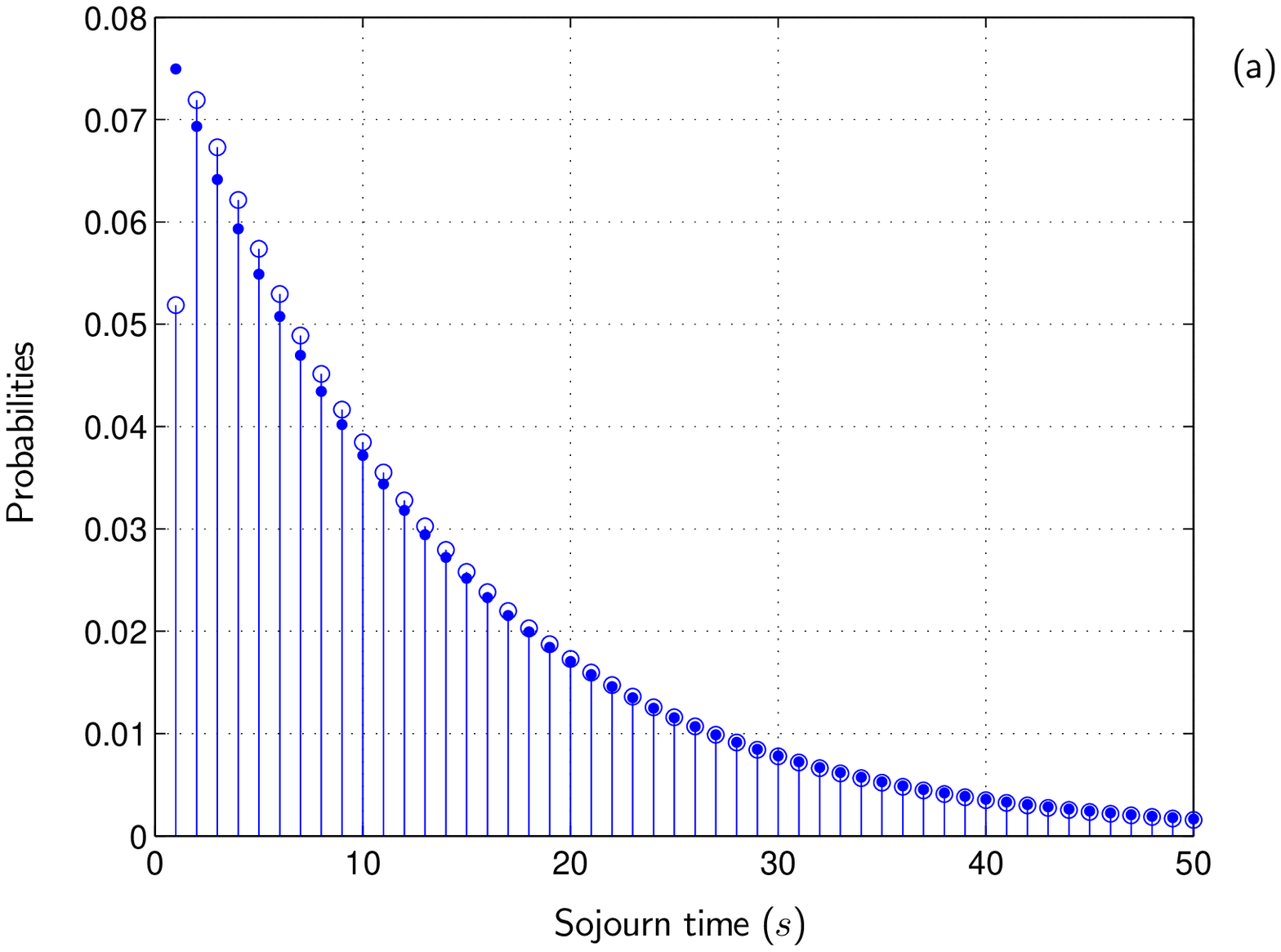} \\
\includegraphics[width=0.8\textwidth]{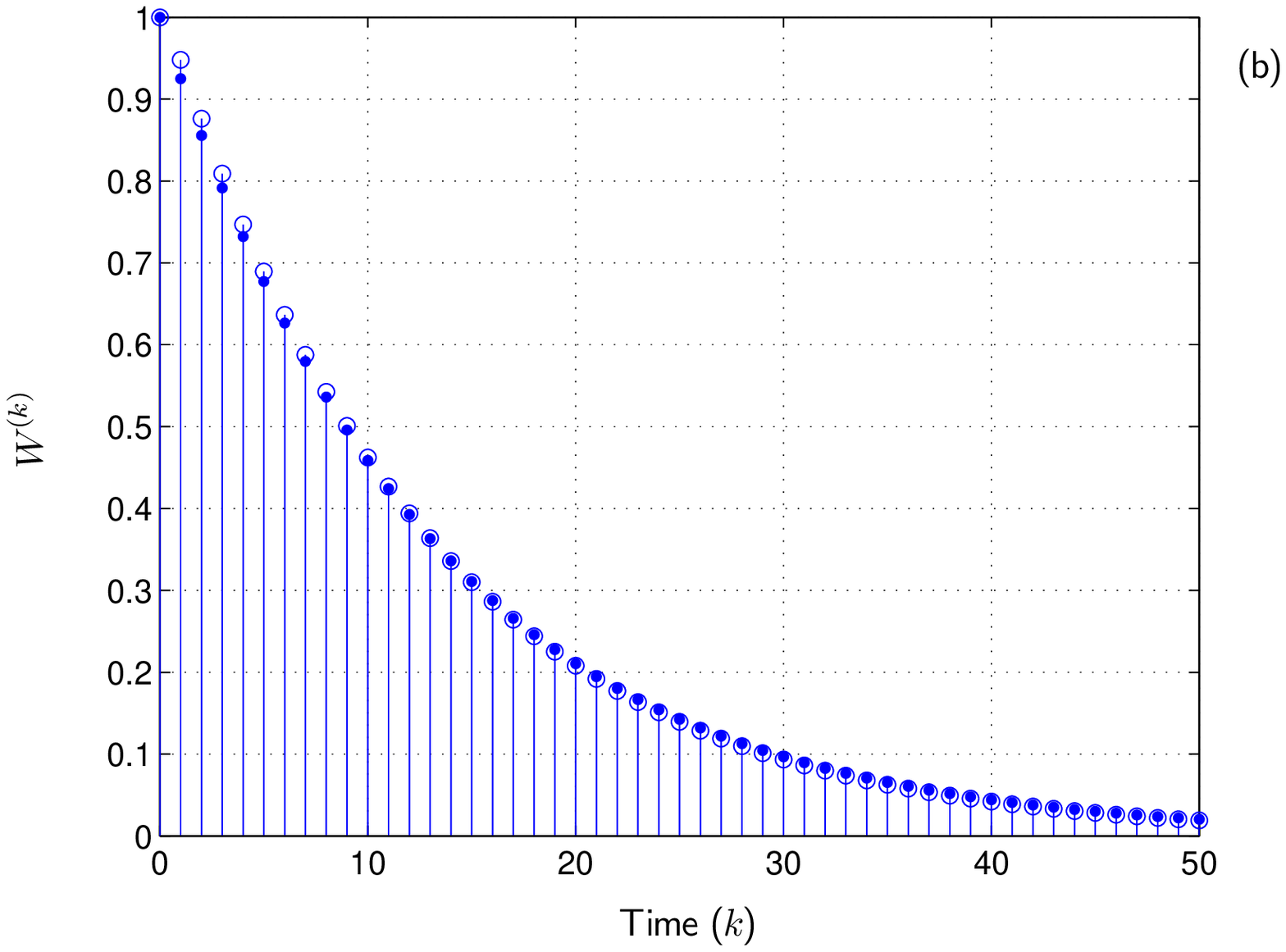} \\
\caption{\textsf{Sojourn time distribution and geometric approximation for basin B1 of Fig.~\ref{figbooex} with $p=0.02$ and $b_{i}^{(0)}=1/6$ $\forall i=1,2,\ldots,6$. (a) Circles: sojourn time distribution $\psi_{s}$ (with mean $\mu=13.3530$); points: approximating geometric distribution $g_{s}$ (with parameter $p_{\mu}=1 / \mu=0.0749$). (b) Circles: $W^{(k)}$; points: geometric approximation $(1-p_{\mu})^{k}$. Maximum deviation $\delta^{*}$ is $2.3007\%$}.}
\label{figmar1}
\end{center}
\end{figure}

\begin{figure}
\begin{center}
\includegraphics[width=1\textwidth]{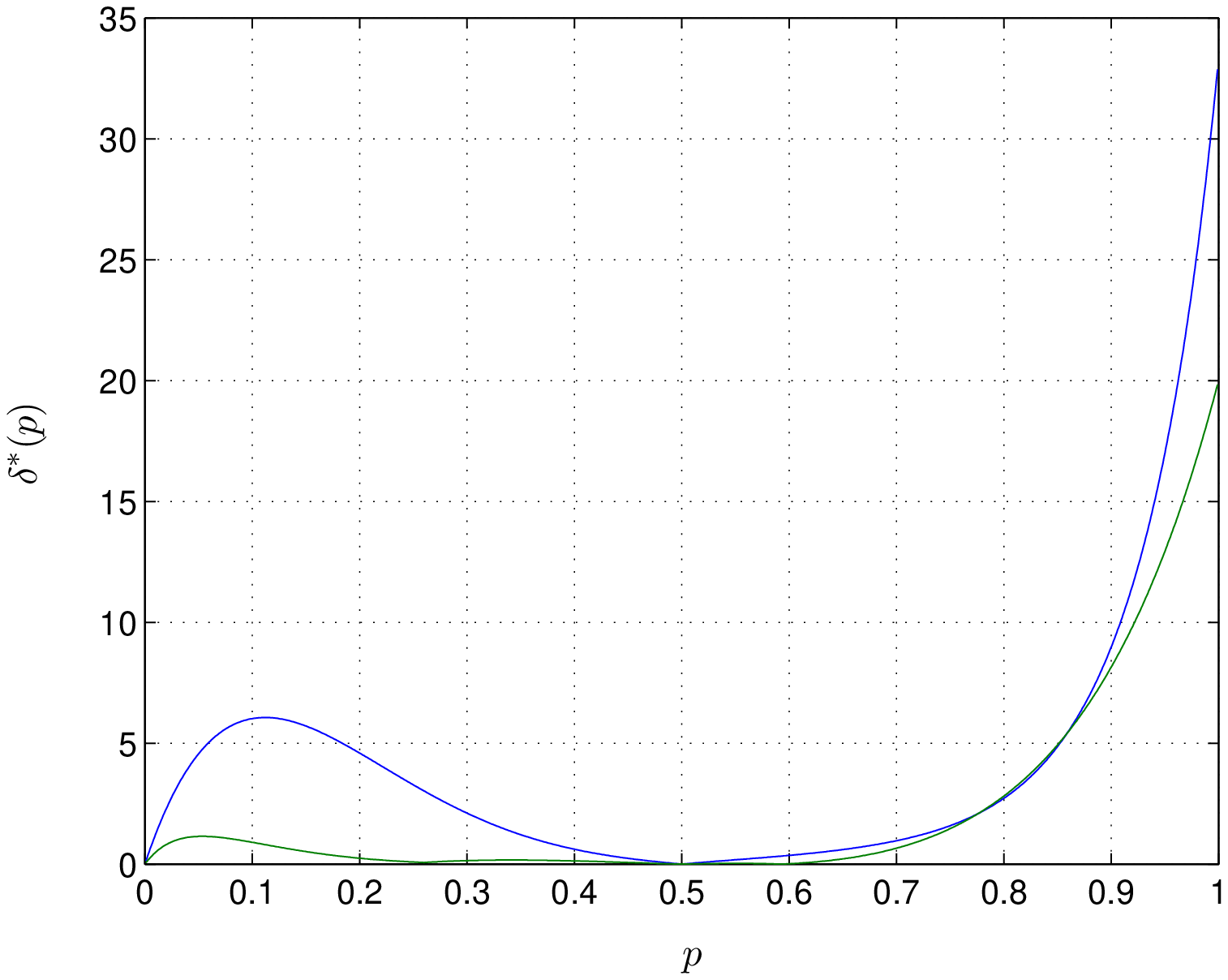}
\caption{\textsf{Maximum deviation $\delta^{*}$ (in $\%$) as a function of $p$ for the two basins of Fig. \ref{figbooex} and initial conditions $b_{i}^{(0)}=1/B$ $\forall i=1,2,\ldots,B$. Blue curve: basin B1. Green curve: basin B2. One has $\delta^{*}(0.5)=0$ for both basins ($\psi_{s}$ is geometric at $p=0.5$)}.}
\label{figdelp}
\end{center}
\end{figure}

\begin{table}
\begin{center}
\begin{tabular}{@{}lcrrrr@{\hspace{1cm}}rrrr@{}}
\toprule
& Ini. cond. & \multicolumn{4}{c}{$\mu$}  &  \multicolumn{4}{c}{$\delta^{*}$} \\ \midrule
\textsf{B1} & $0001$ & 126.00 &  13.52 & 2.37 &   1.54 &   0.39 &   3.44 &   7.47&  10.39\\
& $0101$ & 126.00 &  13.52 & 2.39 &   2.07 &   0.39 &   3.47 &   8.49  & 4.09\\
& $0110$ & 125.50 &  13.02 & 1.94&    2.28 &   0.00 &   0.00  &  0.92 & 10.66\\
& $1001$ & 126.00 &  13.52  & 2.39  &  2.07  &  0.39  &  3.47 &   8.49  &4.09\\
& $\mathbf{1010}$ & 125.50 &  13.02 & 1.94 &  2.28 & 0.00& 0.00& 0.92 & 10.66\\
& $1101$ & 126.00  & 13.52 &   2.37 & 1.54  &  0.39 &   3.44 &   7.47 &  10.39\\ \cmidrule{2-2}
& Mean  & 125.83 &  13.35 & 2.23 & 1.96 &   0.26  &  2.30 & 5.63 & 8.38 \\ \cmidrule{2-10}
& Unif. & 125.83 &  13.35 & 2.23 & 1.96 &  0.26 & 2.30 & 4.59 & 2.72 \\ \midrule
\textsf{B2} & $0000$ & 252.85 &  27.68 &  4.27 &   3.77 &  0.39 &   2.96 &  2.31 &  5.40\\
&$\mathbf{0010}$ & 251.86 &  26.81 &  4.08&  2.25 & 0.00 &  0.11 & 0.50& 20.87\\
&$0011$ & 252.36 &  27.26  &  4.22 &   3.77 &   0.20 &   1.63 &   2.59&  5.42\\
&$0100$ & 251.86 &  26.80 &   4.02  &  3.60 &   0.00  &  0.12 &   1.99 &  0.93\\
&$0111$ & 251.86 &  26.80 &   4.02  &  3.60 &   0.00 &   0.12  &  1.99 &  0.93\\
&$\mathbf{1000}$ & 251.86 &  26.80 &   4.02  &  3.60 &   0.00 &   0.12   & 1.99 &   0.93 \\
&$\mathbf{1011}$ & 251.86 &  26.80 &   4.02  &  3.60  &  0.00  &  0.12  &  1.99  &  0.93 \\
&$1100$ & 252.36 &  27.26  &  4.22  &  3.77 &   0.20 &   1.63  &  2.59&5.42 \\
&$\mathbf{1110}$ & 251.86 &  26.81  &  4.08 &   2.25 &   0.00 &   0.11  & 0.50&  20.87 \\
&$1111$ & 252.85 &  27.68  &  4.27  &  3.77 &   0.39  &  2.96 &   2.31 &   5.40 \\ \cmidrule{2-2}
& Mean & 252.16 &  27.07 & 4.12 & 3.40 & 0.12 &0.99 &1.88 & 6.71 \\ \cmidrule{2-10}
& Unif. & 252.16 &  27.07 & 4.12 & 3.40 & 0.12 & 0.81 &0.24 &2.81 \\ \bottomrule
\end{tabular} 
\caption{\textsf{Mean sojourn time $\mu$ and maximum deviation $\delta^{*}$ (in $\%$) for the two basins of Fig.~\ref{figbooex}. Two types of initial conditions have been considered: (1) the network starts in one state of the basin (successively each state of the basin is taken as initial state) and (2) each element of $\mathbf{b}^{(0)}$ is one divided by the size of the basin (see ``Unif.'' in the Table). The four columns below a parameter ($\mu$ or $\delta^{*}$) correspond from left to right to $p=0.002$, $0.02$, $0.2$ and $0.8$. The mean value for each column is also indicated (see ``Mean''). Attractor states are in bold type}.}
\label{tabdis}
\end{center}
\end{table}
\end{exmp}

\begin{exmp}
In this example, we illustrate the fact that while $p$ is not small, $\psi_{s}$ may be close to a geometric distribution, depending on $\mathbf{b}^{(0)}$. We take basin \textsf{B1} of Fig. \ref{figbooex} and calculate $\psi_{s}$ when initially the network is (a) in state $0101$ and (b) in state $0110$. Fig. \ref{figp2} shows the results. As can be seen in the figure, when the network starts in $0110$, $\psi_{s}$ is close to a geometric distribution which is not the case when the initial state is $0101$.

\begin{figure}
\begin{center}
\includegraphics[width=0.8\textwidth]{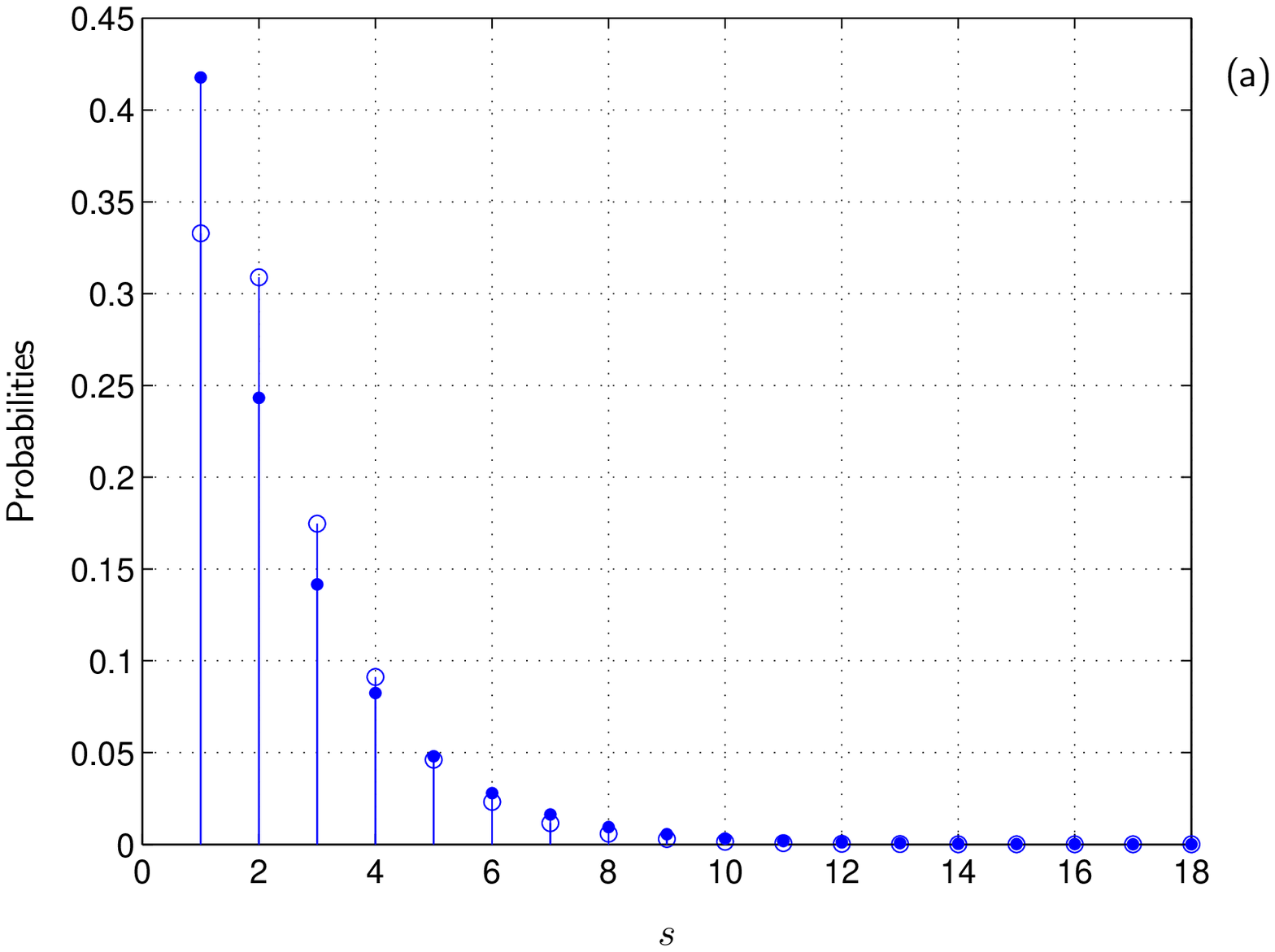} \\
\includegraphics[width=0.8\textwidth]{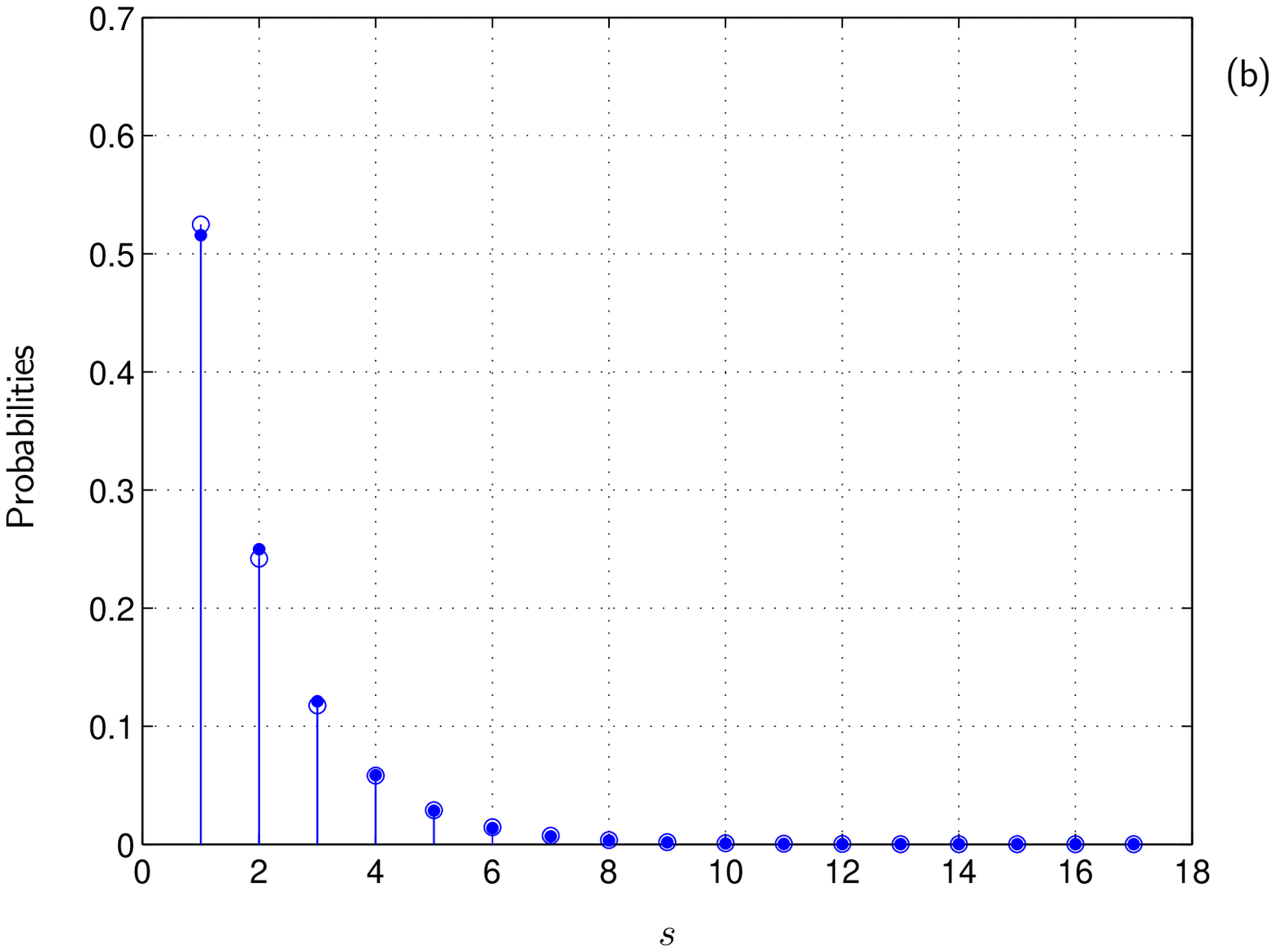} \\
\caption{\textsf{Comparison between two geometric approximations. Two different initial states are taken in basin \textsf{B1} of Fig. \ref{figbooex} with $p=0.2$ (see table~\ref{tabdis}). Circles: probabilities $\psi_{s}$. Points: geometric approximation $g_{s}$. (a) Initial state $0101$. $\delta^{*}=8.49 \%$, $\mu=2.39$. (b) Initial state $0110$. $\delta^{*}=0.92 \%$, $\mu=1.94$}.}
\label{figp2}
\end{center}
\end{figure}
To compare the two approximations, we computed the variances $\sigma^{2}$ and $\sigma^{2}_{g}$ of $\psi_{s}$ and $g_{s}$ respectively and the total variation distance $d_{TV}$ between $\psi_{s}$ and $g_{s}$ (another probability metric) which is given by:

\[
d_{TV}= \frac{1}{2} \sum_{s \geq 1} \vert \psi_{s} - g_{s} \vert.
\] The values of these parameters are given in Table \ref{tabp02}. We see that the total variation distance is about $10$ times greater when the initial state is $0101$ than when it is $0110$. The more geometric $\psi_{s}$, the smaller $d_{TV}$, the better the approximations $\sigma^{2} \approx \sigma^{2}_{g}$ and $\mu \approx \lambda^{*}$.

\begin{table}
\begin{center}
\begin{tabular}{@{}lcc@{}}
\toprule
   & $0101$ & $0110$ \\ \midrule
$\delta^{*}$ (\%)  & 8.49 & 0.92     \\ 
$d_{TV}$  (\%)     & 10.75 & 1.18        \\
$\sigma^{2}$   &  2.38 &  1.91 \\
$\sigma^{2}_{g}$    & 3.34  & 1.82   \\
$\lambda^{*}$  & 2.00 & 2.00  \\
$\mu$ & 2.39 & 1.94 \\
\bottomrule
\end{tabular} 
\caption{\textsf{Comparison between two geometric approximations. See Fig.~\ref{figp2} for details}.}
\label{tabp02}
\end{center}
\end{table}
\end{exmp}

Till now, we have assumed each node of the network has the same probability of being perturbed. In Example $3$ below, we look at how the network of Fig. \ref{figbooex} behaves when one of its nodes is perturbed with a probability which is high compared to the other nodes. Suppose nodes represent proteins. Within the cell interior, some proteins may be more subject to competing reactions than others. These reactions may be assumed to act randomly\footnote{Due to the fluctuating nature of intracellular processes.}, either negatively or positively, on the state of the target proteins. For example, some reactions may lead to protein unfolding while others, like those involving molecular chaperones, may rescue unfolded proteins \citep{DO03}. On the other hand, some proteins may be more sensitive to physico-chemical factors, like temperature or pH, increasing their probability to be perturbed.

\begin{exmp}
In this example, it is assumed that one node is perturbed with a probability which is high compared to the other nodes. The behaviour of the network when $p \to 1$ is complex and will not be discussed here so we take $p_{1}=p_{2}=p_{3}=10^{-3}$ (the first three nodes are rarely perturbed) and $0 < p_{4} \leq 0.9$. Fig. \ref{fighete} shows $\bar{\mathbf{z}}$ for two values of $p_{4}$. For $p_{4}=0.04$ (the blue points) the network behaves as if all the nodes had the same probability $p=0.04$ of being perturbed (see Fig. \ref{figthermo}, the blue points). Simulations have shown this to be true whatever the three rarely perturbed nodes. When $p_{4}$ increases in $]0,0.9]$, the stationary state probabilities do not tend to be equal (see Fig. \ref{fighete}, the case $p_{4}=0.9$), rather, some transient states, typically state $0011$ of \textsf{B2}, tend to be more populated at the expense of attractor states (see attractor states $\mathbf{1000}$, $\mathbf{1010}$, $\mathbf{1011}$ and $\mathbf{1110}$ in Fig. \ref{fighete})\footnote{At the cell population level, this means that a cell population would exhibit phenotypes that could not have been observed (or not easily observed) in the low $p$ regime. Note that these phenotypes are not necessarily survivable for the cells.}. 

\begin{figure}
\begin{center}
\includegraphics[width=1\textwidth]{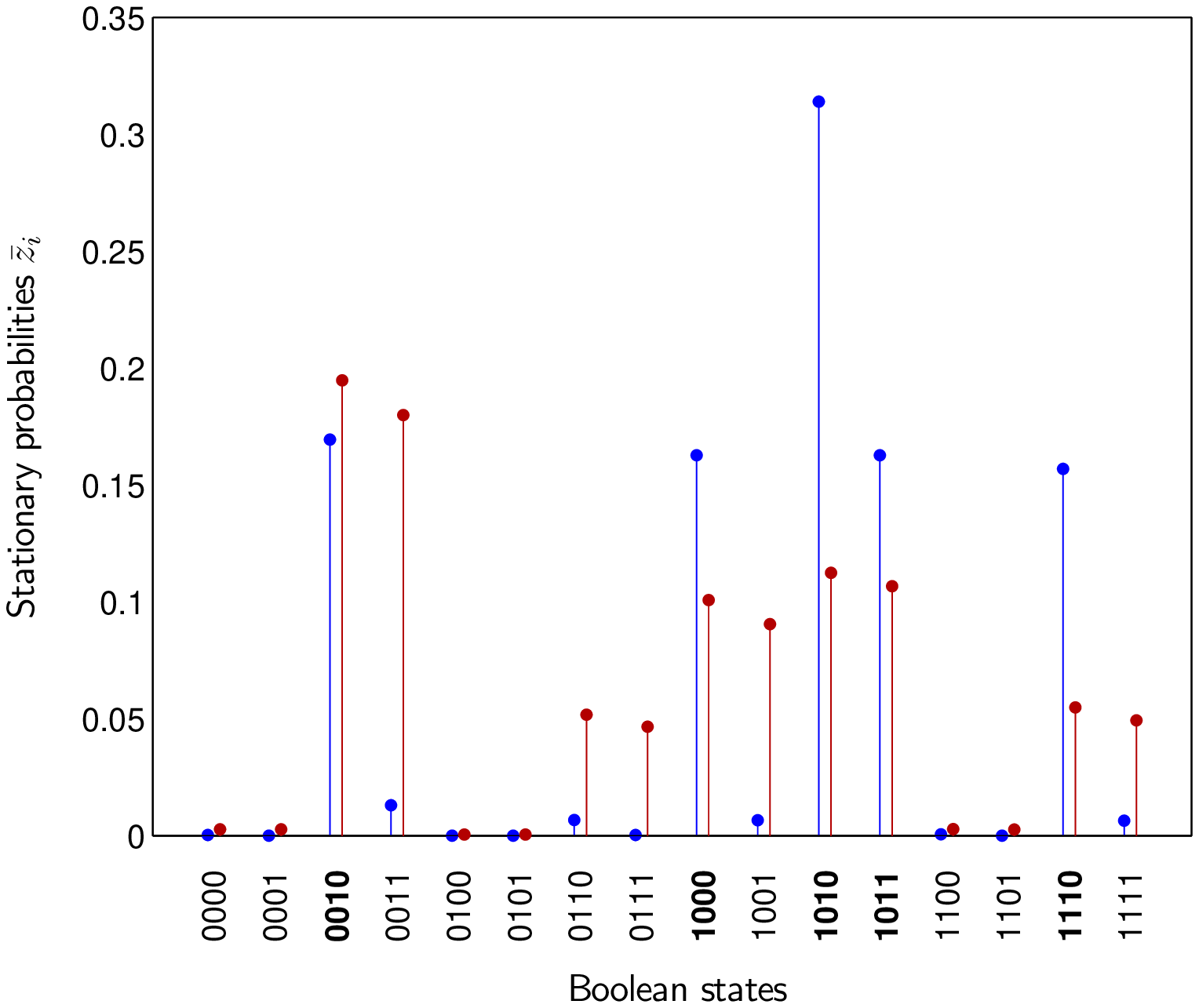}
\caption{\textsf{Stationary state probabilities for the network shown in Fig. \ref{figbooex} and non-equiprobable node perturbations. The first three nodes are rarely perturbed: $p_{1}=p_{2}=p_{3}=10^{-3}$. Blue points: $p_{4}=0.04$; red points: $p_{4}=0.9$. Attractor states are in bold type}.}
\label{fighete}
\end{center}
\end{figure}
Fig. \ref{figmar2}a shows distribution $\psi_{s}$ with approximating distribution $g_{s}$ for basin \textsf{B2}, $p_{4}=0.1$ and state $0011$ as initial state. Maximum deviation $\delta^{*}$ is $8.2011\%$. Note the splitting of $\psi_{s}$ into two subdistributions, one for odd frequencies and the other for even frequencies. Probabilities $W^{(k)}$ and approximations $(1-p_{\mu})^{k}$ are plotted in Fig. \ref{figmar2}b.
\end{exmp}

\begin{figure}
\begin{center}
\includegraphics[width=0.8\textwidth]{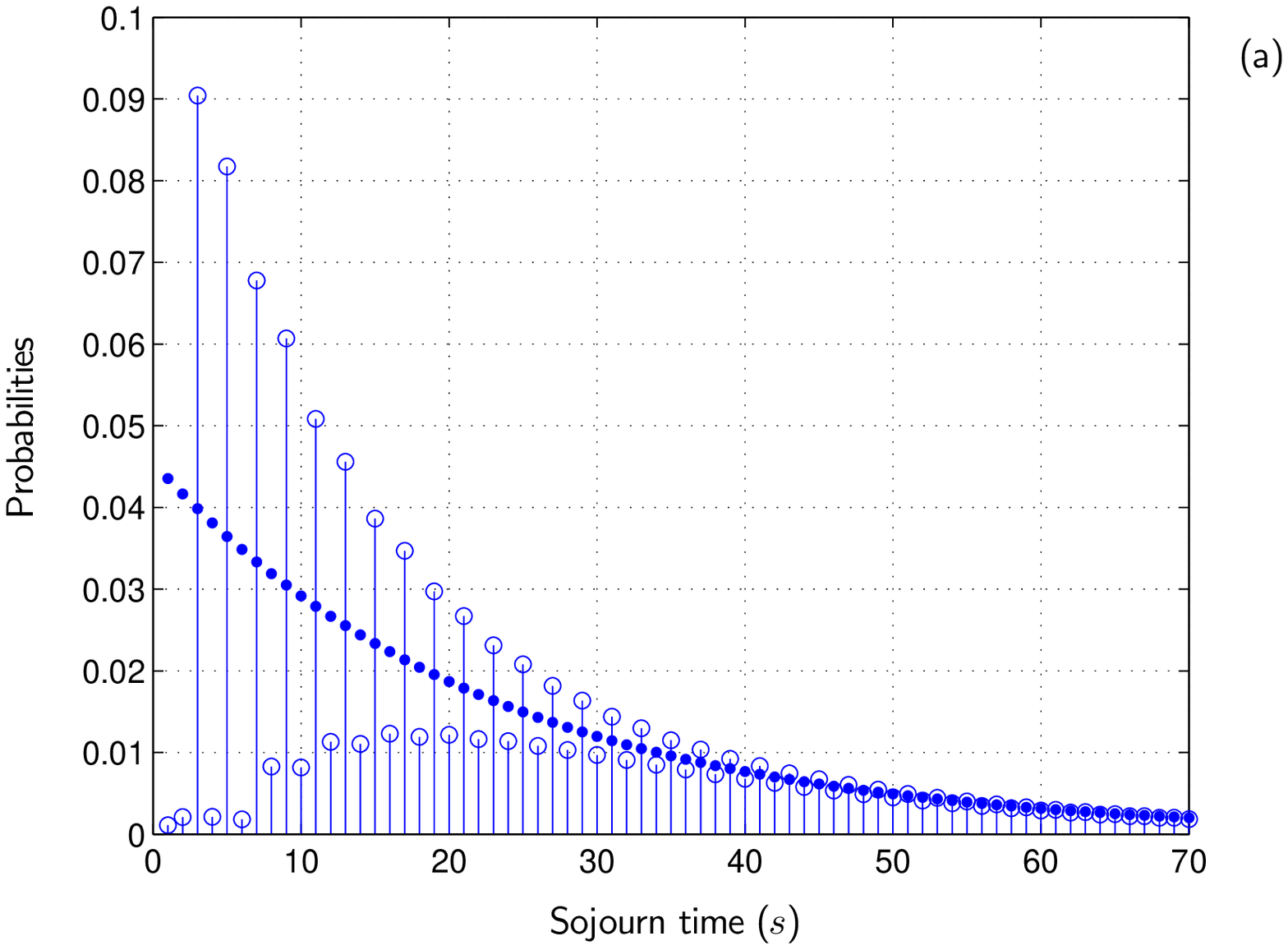} \\
\includegraphics[width=0.8\textwidth]{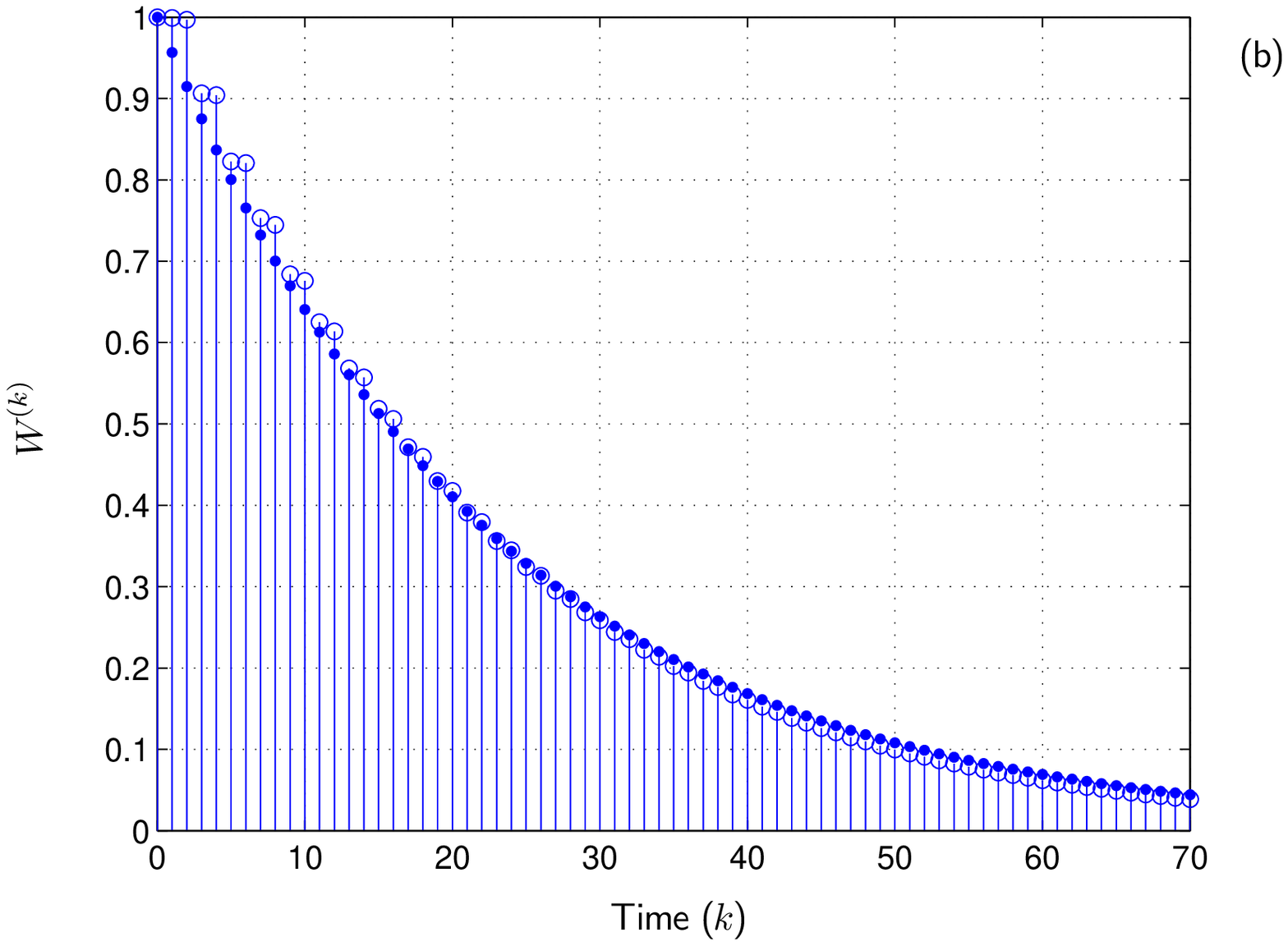} \\
\caption{\textsf{Sojourn time distribution and geometric approximation for basin \textsf{B2} of Fig.~\ref{figbooex}. The first three nodes are rarely perturbed: $p_{1}=p_{2}=p_{3}=10^{-3}$; $p_{4}=0.1$. Initially, the network is in state $0011$. (a) Circles: distribution $\psi_{s}$ (with mean $\mu=22.9653$); points: approximating geometric distribution $g_{s}$ (with parameter $p_{\mu}=0.0435$). For the sake of clarity, sojourn times $s>70$ have been omitted. (b) Circles: $W^{(k)}$; points: geometric approximation $(1-p_{\mu})^{k}$. Maximum deviation $\delta^{*}$ is $8.2011\%$}.}
\label{figmar2}
\end{center}
\end{figure}

\section{The geometric approximation in $(n,C)$ networks}

In Example~1, we illustrated, using the network of Fig.~\ref{figbooex}, the fact that as $p$ tends to $0$, the cdf of the time spent in a basin of attraction approaches the cdf of a geometric (resp. exponential) distribution and that the expected sojourn time tends to be independent of initial conditions (see the decreases of $\delta^{*}$ and variation coefficient as $p$ becomes smaller, respectively in Table \ref{tabdis} and in the text). As will be seen later, for this network, a two-state discrete-time (resp. continuous-time) homogeneous Markov chain may be used for approximating basin transitions in the low $p$ regime, i.e., a coarse-grained description of this network exists in the low $p$ regime. 

Now we address the problem of geometric approximation in randomly constructed $(n,C)$ networks. A random $(n,C)$ network is built by randomly choosing for each node $C$ inputs and one interaction function \citep{KA93}. 

\subsection{$(n,C)$ networks and confidence intervals}

Six $(n,C)$ ensembles were examined taking $n=6$, $8$ or $10$ ($E=64$, $256$ or $1024$) and $C=2$ or $5$. Each basin of a randomly constructed network was perturbed with $p=0.002$, $0.01$, $0.02$, $0.1$, $0.3$, $0.5$ and $0.8$. 

According to the classification of \citet{KA93}, $(n,2)$ networks are complex while $(n,5)$ ones are chaotic. One difference between these two ensembles of networks, which is of particular interest here, is that in the former ``\emph{If the stability of each state cycle attractor is probed by transient reversing of the activity of each element in each state of the state cycle, then for about $80$ to $90$ percent of all such perturbations, the system flows back to the same state cycle. Thus state cycles are inherently stable to most minimal transient perturbations.}'' \citep[p. 201]{KA93}. In chaotic networks, the stability of attractors to minimal perturbations is at best modest \citep[p. 198]{KA93}.

For each $(n,C)$ ensemble, we generated about $2500$ basins (which corresponds to about $700$ to $800$ networks depending on the ensemble). We established confidence intervals for three statistical variables of which two are probabilities:

\begin{enumerate}
\item Consider an $n$-node network. For each of its basins, one can define two conditional probabilities: (1) the conditional probability $\alpha$ of leaving the basin given that one attractor bit out of $nA$ has been flipped, with $A$ the size of the attractor, and (2) the conditional probability $\beta$ of leaving the basin given that one basin bit out of $nB$ has been flipped.

In each basin sample, a small proportion of basins having $\alpha=0$ were found. The $25$th percentile of the relative size $B/E$ of $\alpha=0$ basins was more than $0.8$ for $C=2$ networks and $0.9$ for $C=5$ ones. Thus $\alpha=0$ basins are most often big basins. We calculated $2$ ratios: ratio $\bar{\kappa}$ between the median mean sojourn times of $\alpha=0$ and $\alpha>0$ basins and ratio $\kappa^{*}$ between the maximum mean sojourn times of the two types of basins. These ratios are given in Table~\ref{tabsuper} for $(n,2)$ networks and four $p$ values. In the case of $\alpha=0$ basins, the computation of $\psi_{s}$ for $p=0.01$ and stopping criterion $\hat{\psi}_{s}>0.9999$ varies between a few minutes to several days using a PowerEdge 2950 server with $2$ Quad-Core Intel Xeon processors running at 3.0 GHz. As can be seen from this table, with $p$ decreasing, the maximum mean sojourn time of $\alpha=0$ basins increases drastically compared to that of $\alpha>0$ basins\footnote{For $\alpha=0$ basins, the conditional probability $\alpha_{x}$ of leaving the basin given that $2 \leq x \leq n$ bits of an attractor state have been perturbed may be non null. However $L(p;n,x)=o(p)$ for $2 \leq x \leq n$. On the other hand, for transient states and sufficiently large $k$ one has $\hat{b}_{i}^{(k)} \to 0$ as $p \to 0$.}. The proportion of $\alpha=0$ basins varies in our samples from $1.7$ to $4.1\%$ depending on the ensemble. For fixed $C$, it is a decreasing function of $n$ and for fixed $n$, it is smaller in the chaotic regime than in the complex one. Although we do believe that $\alpha=0$ basins have not to be rejected from a biological interpretation perspective, networks with at least one $\alpha=0$ basin were omitted during the sampling procedure (essentially because of the high computational time that is needed to compute $\psi_{s}$ when $\alpha=0$ and $p$ is small). 

\begin{table}
\begin{center}
\begin{tabular}{@{}llcccc@{}}
\toprule
      &           &    \multicolumn{4}{c}{$p$} \\ \cmidrule(l){3-6}
      &           &  0.03     & 0.05     & 0.1     & 0.5  \\ \midrule
$\bar{\kappa}$ & $n=6$ &  30.1612  &  19.9012 & 12.1390 &    6.8534 \\
          & $n=8$ &  44.3541  &  26.0081 & 15.1438 &   11.9929 \\
          & $n=10$&  35.5351  &  21.6376 & 14.5090 &   10.3930 \\

$\kappa^{*}$ & $n=6$ &  122.3414  & 19.3446  &  3.6270  &  1.0000 \\
          & $n=8$ &  300.3482  & 32.4353  &  3.1643  &  1.0000 \\
          & $n=10$&  542.8109  & 67.3100  &  4.9887  &  1.0000 \\
\bottomrule
\end{tabular} 
\caption{\textsf{Comparison between median mean sojourn times of $\alpha=0$ and $\alpha>0$ basins (ratio $\bar{\kappa}$) and between maximum mean sojourn times of both types of basins (ratio $\kappa^{*}$) for $(n,2)$ networks and four $p$ values. Calculation of $\psi_{s}$: at time $0$, the states of the basin are equiprobable. The proportions of $\alpha=0$ basins for samples $n=6$, $8$ and $10$ were found to be $4.0759$, $3.1989$ and $2.5971\%$ respectively}.}
\label{tabsuper}
\end{center}
\end{table}

\begin{figure} 
\begin{center}
\includegraphics[width=1\textwidth]{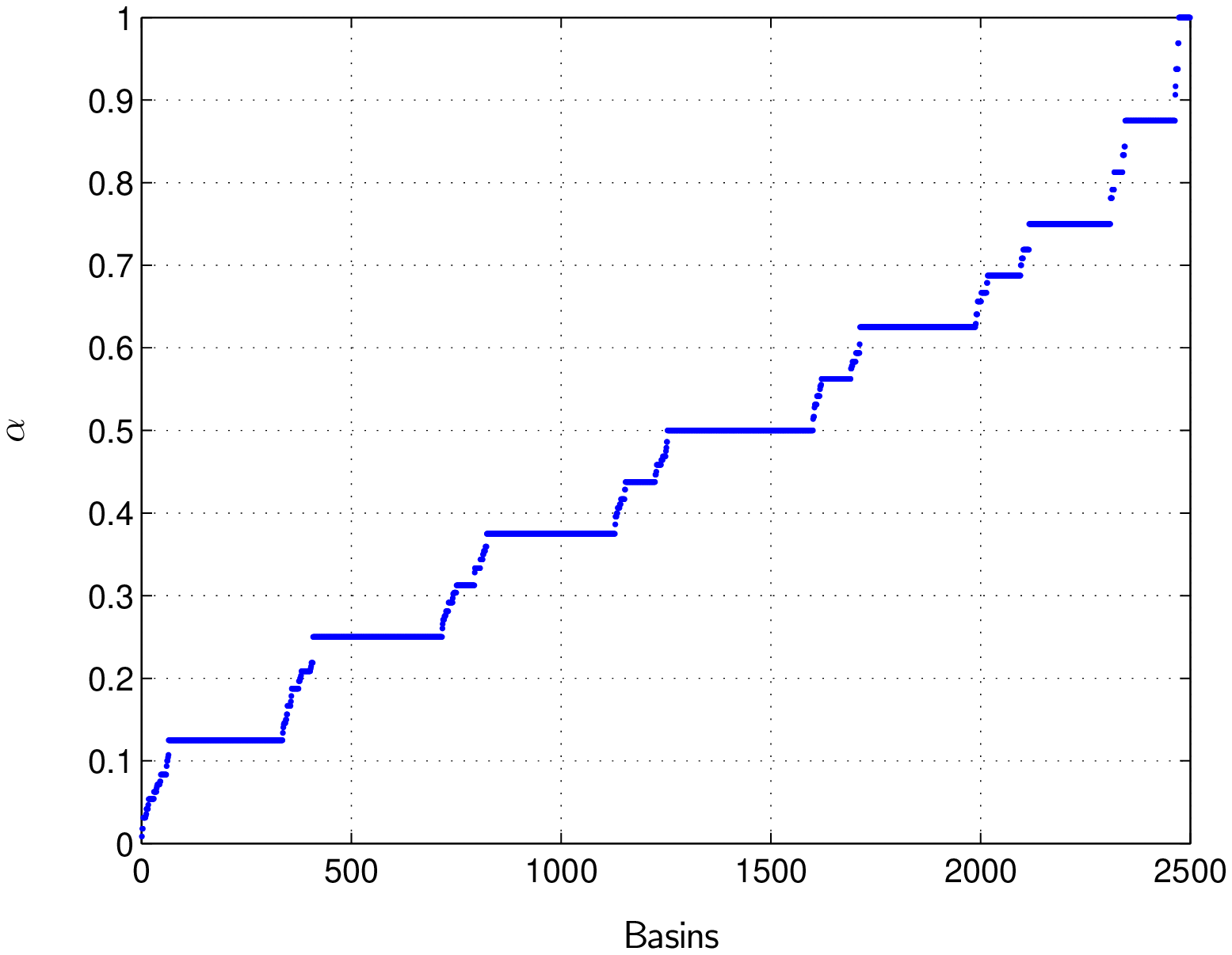}
\caption{\textsf{Conditional probabilities $\alpha>0$ sorted in ascending order, sample $(8,2)$. About $74\%$ of the data points are in main probability levels. These correspond to probabilities $1/8,2/8,\ldots,1$}.}
\label{quanti}
\end{center}
\end{figure}

Fig. \ref{quanti} shows the conditional probabilities $\alpha$ for sample $(8,2)$. For an $n$-node network, one can define $n$ main probability levels $1,2,\ldots,n$ corresponding to probabilities $1/n,2/n,\ldots,1$. The percentage of data points located in main probability levels are given in Table~\ref{tabmpl} for the six samples and the two conditional probabilities $\alpha$ and $\beta$. It can be seen from this table that whatever the connectivity and the conditional probability, this percentage decreases when $n$ increases. For a given conditional probability and any $n$, it is greater in the complex regime than in the chaotic one. 

The statistical analysis of the six data samples has shown the following. For $C=2$, there is a clear quantization of $\alpha$ and a weak one of $\beta$. Also the histograms of $\alpha$ and $\beta$ are symmetric (skewness equal to $\sim 0.2$). For $C=5$, the quantization of $\alpha$ is weak and there is no obvious quantization of $\beta$. The histograms of $\alpha$ and $\beta$ are negatively skewed (skewness equal to $\sim -0.9$) and the last two main probability levels $(n-1)$ and $n$ are strongly populated compared to the other ones.

\begin{table}
\begin{center}
\begin{tabular}{@{}ccccc@{}}
\toprule
$C$   & $n$ &  & $\alpha$ & $\beta$ \\ \midrule
$2$   & $6$ &   &  78.6 & 52.0 \\
      & $8$ &   &  73.9 & 45.9 \\
      & $10$&   &  70.2 & 35.3 \\ \midrule
$5$   & $6$ &   &  56.6 & 31.5 \\
      & $8$ &   &  50.8 & 23.5 \\
      & $10$&   &  43.6 & 15.4 \\
\bottomrule
\end{tabular} 
\caption{\textsf{Percentages of data points located in main probability levels for the six ensembles $(n,C)$ and the two conditional probabilities $\alpha>0$ and $\beta>0$}.}
\label{tabmpl}
\end{center}
\end{table}

Table \ref{tabalga} gives the mean and median of probabilities $\alpha$ and $\beta$ for the six ensembles $(n,C)$. For fixed $n$, the mean and median of both conditional probabilities are higher for chaotic basins than for complex ones. Additionally, for $C=2$ basins, the mean and median decrease with $n$ while for $C=5$, both increase with $n$ (except for the median of $\alpha$). 

\begin{table}
\begin{center}
\begin{tabular}{@{}cccccc@{}}
\toprule
 & & \multicolumn{2}{c}{$\alpha$} & \multicolumn{2}{c}{$\beta$} \\ \cmidrule(l){3-6}
$C$   & $n$ &  Mean & Median & Mean & Median \\ \midrule
$2$   & $6$    & 0.5057  & 0.5000  & 0.4817  & 0.5000 \\
      & $8$    & 0.4523  & 0.4375  & 0.4294  & 0.4144 \\
      & $10$   & 0.4149  & 0.4000  &  0.3907 & 0.4000 \\ \midrule
$5$   & $6$    & 0.6869    & 0.7917    & 0.6852    & 0.7778 \\
      & $8$    & 0.7032    & 0.8125    & 0.7038    & 0.8219 \\
      & $10$   & 0.7050    & 0.8125    & 0.7106    & 0.8375 \\
\bottomrule
\end{tabular} 
\caption{\textsf{Mean and median of conditional probabilities $\alpha>0$ and $\beta>0$ for the six ensembles $(n,C)$}.}
\label{tabalga}
\end{center}
\end{table}

\item The third statistical variable is the mean time spent in a basin, $\mu$. For $C=2$ and any $n$, there is a clear quantization at $p=0.002$ which rapidly disappears as $p$ increases. For $C=5$ and any $n$, there are only two levels of quantization at $p=0.002$ and these rapidly vanish as $p$ increases. Also notice that the minimum of $\mu$ is equal to the mean specific path $d=\tau$ which, according to (\ref{eqmoygeo}), does not depend on $C$. 
\end{enumerate}

Most of the limits of the $95\%$ confidence intervals ranged between $1.5$ and $2.5\%$ (for $\alpha$ and $\beta$: confidence interval for the mean if $C=2$, for the median and for the trimmed mean if $C=5$; for $\mu$: confidence interval for the trimmed mean)\footnote{Two methods were used: a nonparametric method based on the binomial distribution (for the median only) and the \emph{bootstrap} method (used for the median and the trimmed mean). For the trimmed mean, we averaged the sample data that were (1) between the $25$th and $75$th percentiles, (2) less than the $75$th percentile and (3) less than the $90$th percentile.}. 

\subsection{Simulation results and discussion}

Most of the basin variables (such as $\mu$ or $\delta^{*}$) were positively skewed. Therefore for each of these variables we calculated quartiles $Q_{1}$, $Q_{2}$ (the median) and $Q_{3}$. 

For the calculation of $\mu$, two types of initial conditions were considered:

\begin{enumerate}
\item Each element of $\mathbf{b}^{(0)}$ is equal to $1/B$. We call this condition the uniform initial condition. 
\item We pick one state of the basin at random and place initially the network in that state (the elements of $\mathbf{b}^{(0)}$ are all $0$ except one which is equal to $1$). We call this condition the random initial condition. 
\end{enumerate}

\begin{figure}
\begin{center}
\includegraphics[width=1\textwidth]{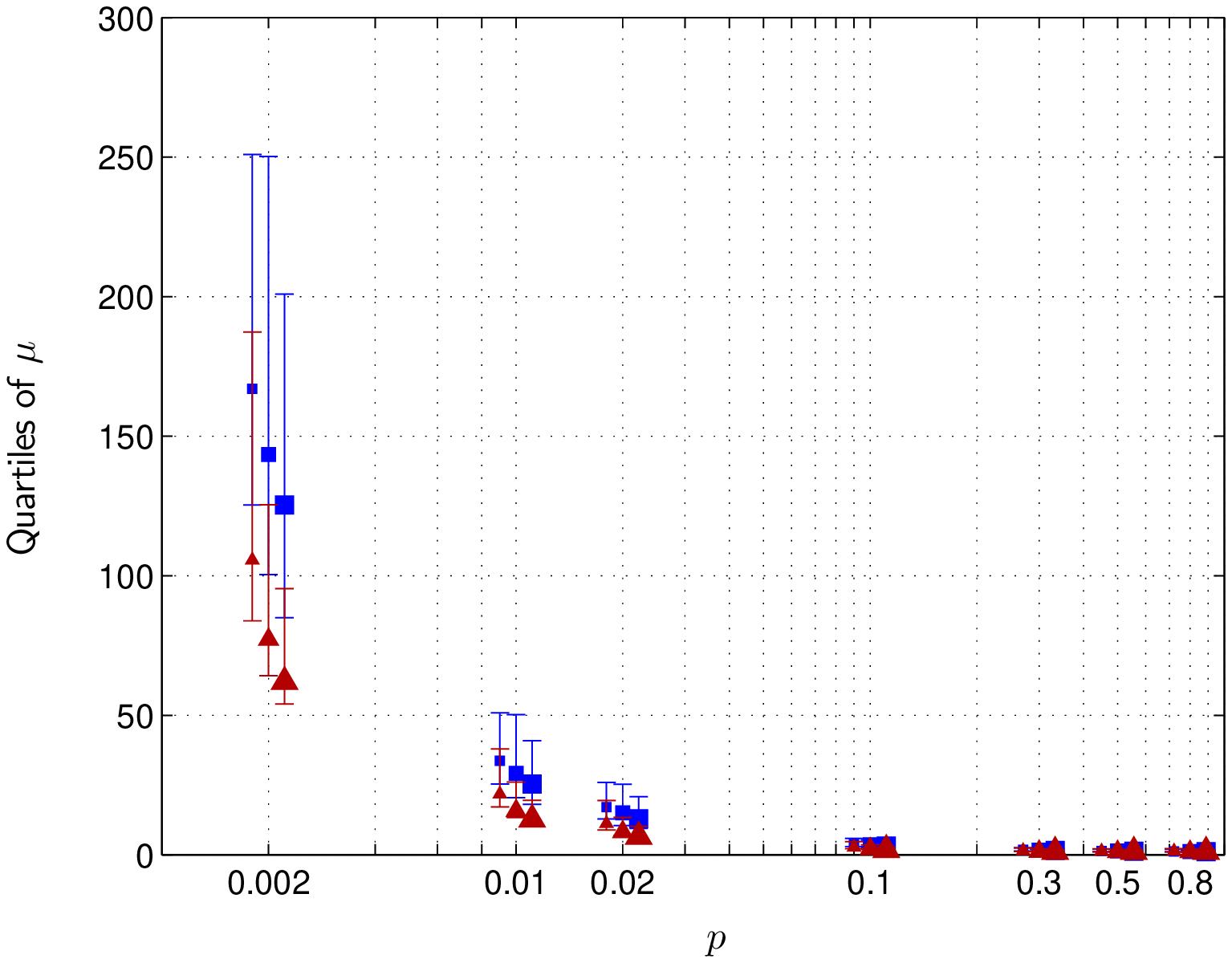}
\caption{\textsf{Median and interquartile range of mean sojourn time $\mu$ versus $\ln p$ for the six ensembles $(n,C)$. Calculation of $\psi_{s}$ with the uniform initial condition. Quartiles $Q_{1}$ and $Q_{3}$ are indicated by horizontal bars. Blue squares: complex regime ($C=2$); red triangles: chaotic regime ($C=5$). The size of a symbol (square or triangle) is proportional to $n$}.}
\label{figus1}
\end{center}
\end{figure}

Let's start with the basin variable $\mu$. First the uniform initial condition. Coefficients of skewness for $\mu$ distributions mostly ranged from $5$ to $15$ with mean of $10.0568$ (strong skewness). For example, for sample $(8,5)$ with $p=0.01$, we obtained a coefficient of skewness of $10.2415$, a mean of $44.7471$, a median of $15.7922$, a $75$th percentile of $26.0523$, a $99$th percentile of $608.4743$ and a maximum of $3.0117 \times 10^{3}$. A small proportion of the mean sojourn times are therefore very far from the median (taken here as the central tendency). The three quartiles of $\mu$ versus $\ln p$ are shown in Fig.~\ref{figus1} for the six ensembles $(n,C)$. The blue squares correspond to $C=2$ and the red triangles to $C=5$. The size of a symbol is proportional to $n$. For the sake of clarity, the absciss{\ae} of the points corresponding to ensembles $(6,C)$ and $(10,C)$ have been translated (respectively to the left and to the right of the $p$ values). 

It can be seen from Fig. \ref{figus1} that: (1) for fixed $p$ and any $C$, the median of $\mu$ is a decreasing function of the size $n$ of the network (not true at $p=0.5$ and $0.8$), although $\alpha$ decreases with $n$ (see Table~\ref{tabalga}). (2) For fixed $C$ and any $n$, the median of $\mu$ is a decreasing function of $p$. (3) For fixed $p$ and any $n$, median $\mu$ of chaotic basins ($C=5$) is less than that of complex ones ($C=2$). At $p=0.002$, the medians of $\mu$ for $C=5$ ensembles are respectively $1.58$, $1.86$ and $2.02$ times smaller than those for $C=2$ ones. Similar but smaller values were found for $p=0.01$ and $p=0.02$. We therefore confirm the results of \citet[p. 198, 201 and 488-491]{KA93} that chaotic basins are less stable to node perturbations than complex ones.

\begin{figure}
\begin{center}
\includegraphics[width=0.8\textwidth]{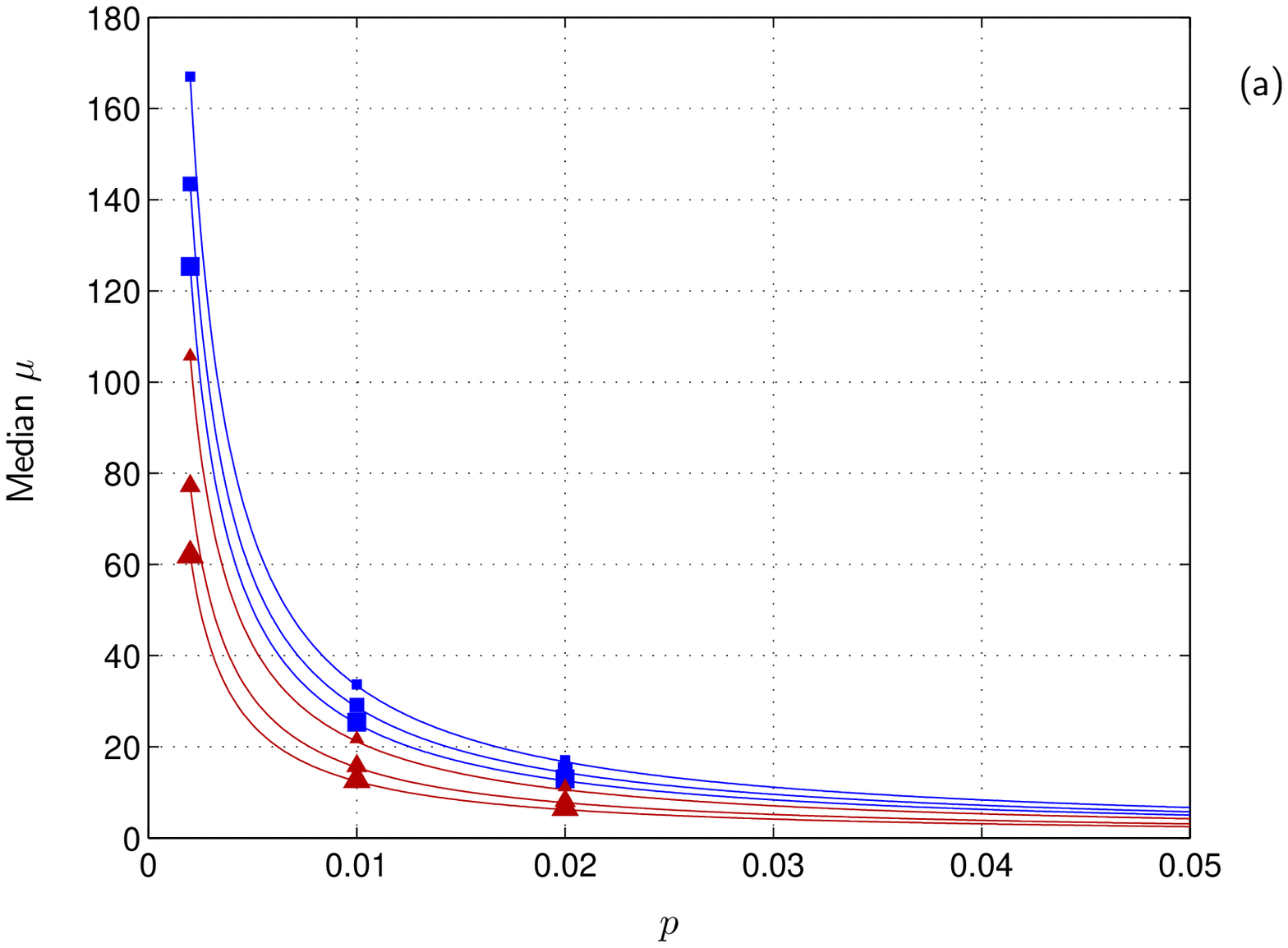} \\
\includegraphics[width=0.8\textwidth]{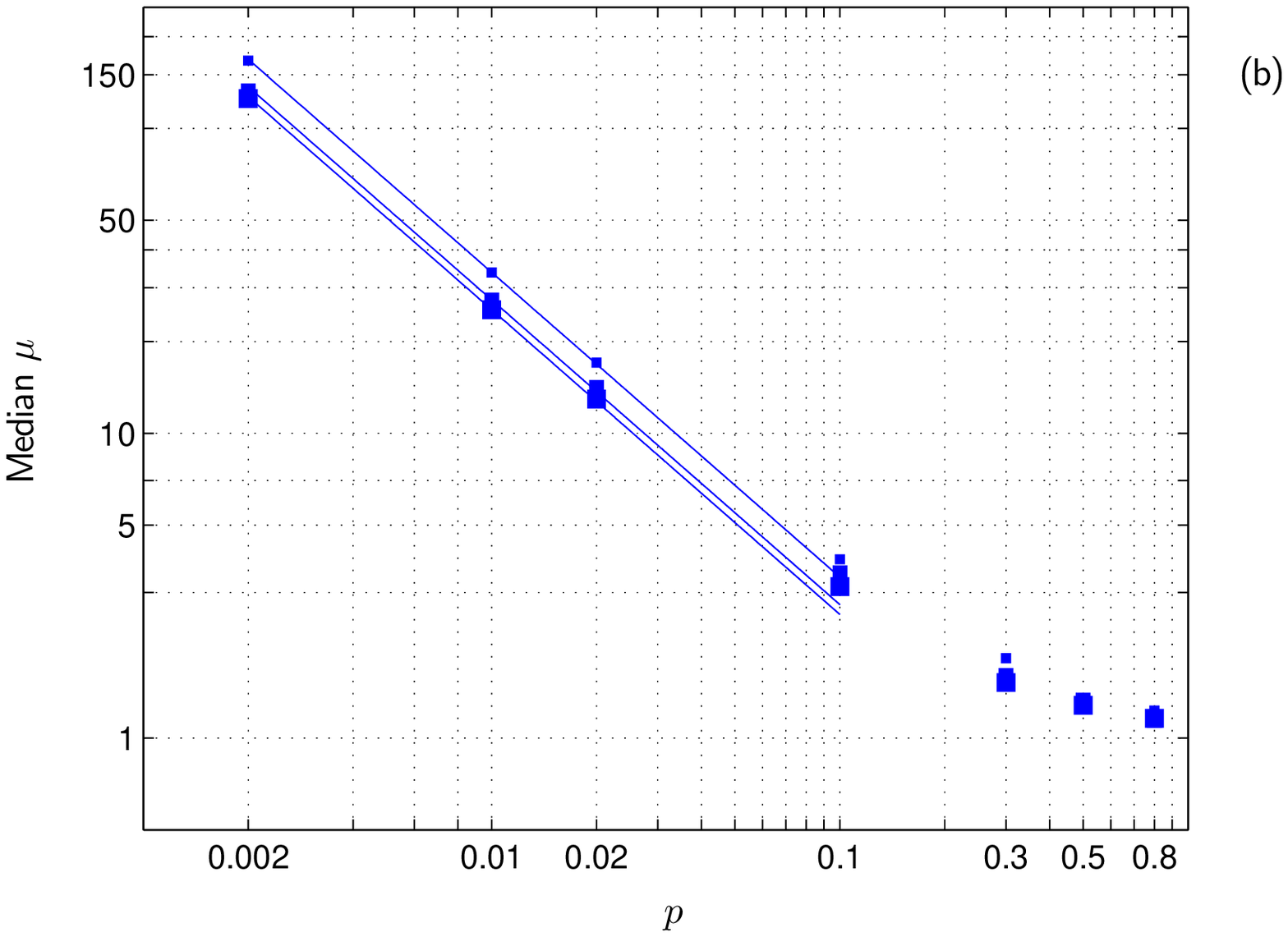} \\
\caption{\textsf{Fit of (\ref{eqhyp}) by the least squares technique. (a) Linear scale for both axis, fits for the six ensembles $(n,C)$. For the sake of clarity, the points with $p>0.05$ have been omitted. Blue squares: $C=2$; red triangles: $C=5$. The size of a symbol is proportional to $n$. (b) Logarithmic scale for both axis (ln-ln), fits for ensembles $(n,2)$ only. The size of a symbol is proportional to $n$}.}
\label{figus2}
\end{center}
\end{figure}

We looked for the relationship between the median of $\mu$ and $p$. We found that for sufficiently small $p$:

\begin{equation}\label{eqhyp}
Q_{(2;\mu)} \approx \frac{c_{2}}{p},
\end{equation} i.e. the median of the mean sojourn time is inversely proportional to $p$. The proportionality constant $c_{2}$ has been estimated by the least squares method for the six ensembles $(n,C)$ and was found to decrease when $n$ increases (only the first three data points were fitted, i.e. the points with abscissa $p=0.002$, $0.01$ and $0.02$). For $C=2$, we obtained $c_{2}=0.3342$, $0.2871$ and $0.2510$; for $C=5$, $c_{2}=0.2117$, $0.1547$ and $0.1242$. Thus for fixed $n$, chaotic basins are less sensitive to a variation in $p$ than complex ones. The result of the least squares fit is presented in Fig. \ref{figus2}. Graph (a) shows the hyperbolic relationship between the median of $\mu$ and $p$ for the six ensembles $(n,C)$. For the sake of clarity, the functions were drawn up to $p=0.05$. When a logarithmic scale for each axis is used, one obtains the graph (b) which shows for the three ensembles $(n,2)$ a linear relationship between $\ln Q_{(2;\mu)}$ and $\ln p$ at low $p$ (up to $p=0.02$ on the graph)\footnote{The three straight lines in Fig.~\ref{figus2}b were obtained by the least squares method applied to the linearized problem: $Y_{2}=a_{2}-X$ where $Y_{2}=\ln Q_{(2;\mu)}$, $a_{2}=\ln c_{2}$ and $X=\ln p$. Only the first three data points were fitted.}. The same rule applies to the three ensembles $(n,5)$. We will see further how to express $c_{2}$ in function of $n$ and median $\alpha$ probability.

With the random initial condition, the quartiles of $\mu$ are very close to those obtained with the uniform initial condition. 

We then calculated for each basin the relative error between the mean sojourn time obtained from the uniform initial condition and that obtained from the random initial condition. This error is null at $p=0.5$ (as for the network of Fig. \ref{figbooex}, this is because $\psi_{s}$ is geometric at $p=0.5$). The quartiles of the error tend to $0$ as $p \to 0$ and they reach a maximum at $p=0.1$ whatever the network ensemble. Thus the smaller $p$, the more $\mu$ is independent of initial conditions.

\begin{figure}
\begin{center}
\includegraphics[width=1\textwidth]{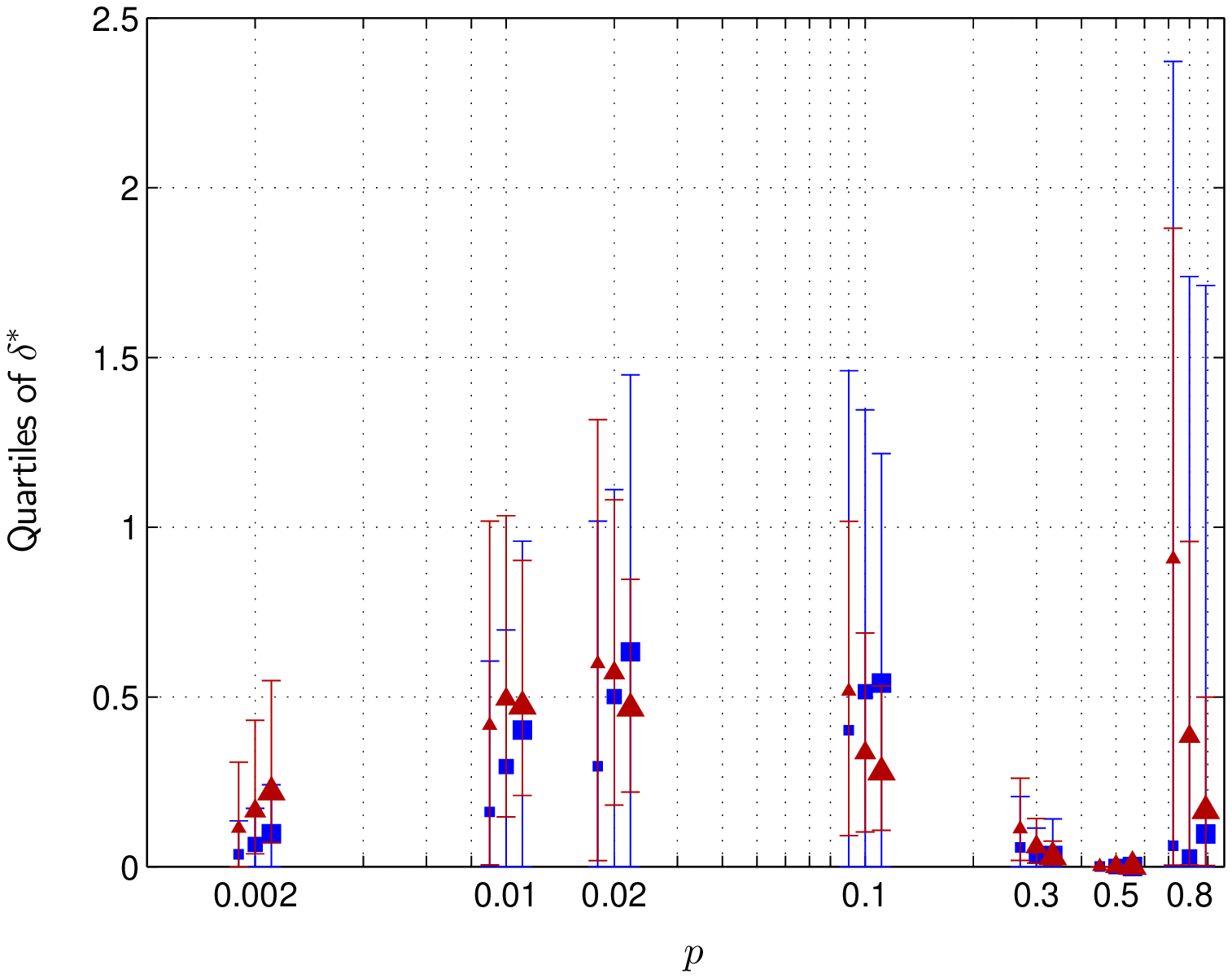}
\caption{\textsf{Geometric approximation for $(n,C)$ ensembles. Median and interquartile range of $\delta^{*}$ (in $\%$) versus $\ln p$. Calculation of $\psi_{s}$ with the uniform initial condition. Blue squares: $C=2$; red triangles: $C=5$. As $p \to 0$, the three quartiles of $\delta^{*}$ tend to $0$}.}
\label{figdel}
\end{center}
\end{figure}

To end, we turn to the maximum deviation $\delta^{*}$. First the uniform initial condition. The results for the six $(n,C)$ ensembles are presented in Fig. \ref{figdel}. At $p=0.002$ and for fixed $C$, the quartiles increase linearly with $n$, except the first quartile of $C=2$ basins which is constant and approximately equal to $0.0005 \%$ (whatever $n$, $25\%$ of $C=2$ basins have a sojourn time that is geometric or closely follows a geometric distribution). 

For both connectivities, the functions $Q_{(2;\delta^{*})}(p)$ and $Q_{(3;\delta^{*})}(p)$ have similarities with the functions $\delta^{*}(p)$ of Fig. \ref{figdelp}: they tend to $0$ when $p \to 0$, they have at least one local maximum in $0 < p < 0.5$ and are null at $p=0.5$. 

With the random initial condition, the second and third quartiles of $\delta^{*}$ increase for most of the six ensembles $(n,C)$ compared to the uniform case. Qualitatively, the behaviour of the quartiles with respect to $p$ is similar to the one found with the uniform initial condition. 

To summarize sections 3 and 4, we have the following proposition: 

\begin{prop}
Consider a basin of a noisy Boolean network. As $p$ tends to $0$ then:
\begin{enumerate}
\item whatever the initial conditions, $\mu \to \infty$.
\item $\mu$ tends to be independent of the initial conditions: $\mu \to \lambda^{*}$ with $\lambda^{*}>1$ the Perron-Frobenius eigenvalue of fundamental matrix $(I-Q)^{-1}$.
\item whatever the initial conditions, $\delta^{*} \to 0$, where $\delta^{*}$ is the Kolmogorov distance between the cumulative distribution function of the sojourn time $S$ in the basin and that of a geometric distribution $g_{s}$ having the same mean as $S$. 
\item $g_{s}$ converges to an exponential distribution.
\end{enumerate}
\end{prop}

Proposition~3 does not guarantee the existence of a coarser representation of a \textsf{NBN} in the low $p$ regime. What can be said from this proposition is that if such a representation exists, then it is in general an approximation of the original process and it can always be expressed in a discrete or continuous time framework. Thus we are lead to the proposition below that will be discussed in more details in the next section:

\begin{prop}
Consider a noisy Boolean network and suppose there exists a coarser time-homogeneous representation of this network in the low $p$ regime. Then the dynamics between the basins of the network may be approximated by a system of linear ordinary differential equations with size being equal to the number of attractors of the network.
\end{prop}

Recall that Proposition~3 does not mean that $\psi_{s}$ can never be geometric nor be approximated by a geometric distribution when $p$ is not sufficiently close to $0$. For example we know $\psi_{s}$ is geometric when $p=0.5$ or when $B=1$ whatever $0 < p < 1$. One difference between a strongly and a weakly perturbed network is that in the latter, since the mean specific path $d$ is large compared to $1$, the network spends long periods on the attractors without being perturbed. Under normal conditions, biochemical networks are supposed to work in the low $p$ regime because as explained in 1.2, this regime is associated with functional stability.

\subsection{Approximation formula for the mean time spent in a basin of a \textsf{NBN}}

For sufficiently small $p$, $(i,j)$th element of $\Pi'$ can be approximated by ($i \neq j$):

\[
\pi_{ij}' = p^{h_{ij}}q^{n-h_{ij}} \approx \left\{ \begin{array}{ll} 
p & \textrm{if} \quad h_{ij}=1, \\
0 &  \textrm{if} \quad h_{ij}>1.
\end{array} \right.
\] From (\ref{eqps13}) then we may write:

\begin{equation}\label{eqps14}
p_{e}(k,k+1) \approx p \sum_{i \in \textsf{B}} \Gamma_{i}^{1} \hat{b}_{i}^{(k)}.
\end{equation} We see that in the low $p$ regime, the probability $p_{e}(k,k+1)$ depends on time and initial conditions. 

Now for sufficiently small $p$, we have the following:

\begin{equation}\label{eqmue2}
\mu \approx 1 / n p \alpha,
\end{equation} which, from Proposition 2, is equivalent to:

\[ 
\lim_{p \to 0} n p \alpha \lambda^{*} = 1.
\] To show approximation formula (\ref{eqmue2}), we consider two cases:

\begin{enumerate}
\item Suppose for any given $0 < p < 1$, the elements of vector $\mathbf{a}$ are equal. Then from (\ref{eqps13}) $p_{e}(k,k+1)$ is constant which means $\psi_{s}$ geometric. Thus taking $p$ sufficiently small, $\Gamma_{i}^{1}$ must be the same for all state $i \in \textsf{B}$ and therefore from (\ref{eqps14}):

\begin{equation}\label{eqmalp}
1/ \mu = p_{\mu} \approx p \Gamma^{1}=n p \alpha,
\end{equation} with $\alpha= A \Gamma^{1} / n A$ and $A$ the size of the attractor. 

Note that: (1) if $A=B=1$, it comes that $p_{\mu}=r \approx np$ and therefore $\mu = \tau \approx 1 / n p$. (2) If $A=B=2$, then $\psi_{s}$ is geometric since in this case $Q$ is a symmetric matrix of size $2$. If one starts with probability $1$ in one of the two states, then each probability $\hat{b}_{i}^{(k)}$ will be periodic with period $2$. (3) If $\psi_{s}$ is geometric then the parameter of the distribution may be written:
\[
p_{\mu}=\sum_{x=1}^{n} L(p;n,x) \alpha_{x},
\] with $\alpha_{1}=\alpha$ and $\alpha_{2},\alpha_{3},\ldots,\alpha_{n}$ conditional probabilities defined as $\alpha$ except that rather than perturbing one bit we perturb $2,3,\ldots,n$ bits simultaneously. Taking $p$ sufficiently small in this formula, we retrieve approximation (\ref{eqmalp}).
\item If $\psi_{s}$ is not geometric, then for sufficiently small $p$ it comes from (\ref{eqps14}) that $\forall \mathbf{b}^{(0)}$ $p_{e}(k,k+1)$ must tend to $p \sum_{i \in \textsf{A}} \Gamma_{i}^{1} / A$ as $k \to \infty$ since $\hat{b}_{i}^{(k)}$ must tend to a value which is close to $0$ $\forall i \in \textsf{B} \setminus \textsf{A}$ and to a value which is close to $1/A$ $\forall i \in \textsf{A}$. From Propositions 1 and 2 then, we get (\ref{eqmue2}) with:

\[
\alpha=\sum_{i \in \textsf{A}} \Gamma_{i}^{1} / n A.
\] 
\end{enumerate}

\begin{rem}
We see from (\ref{eqmue2}) that the quantization of $\mu$ observed at $p=0.002$ for $(n,2)$ ensembles (and to a lesser extent for chaotic ensembles) is a direct consequence of the quantization of $\alpha$ for these ensembles (see Fig. \ref{quanti} and point 2 of 4.1). 
\end{rem}

\begin{figure}
\begin{center}
\includegraphics[width=1\textwidth]{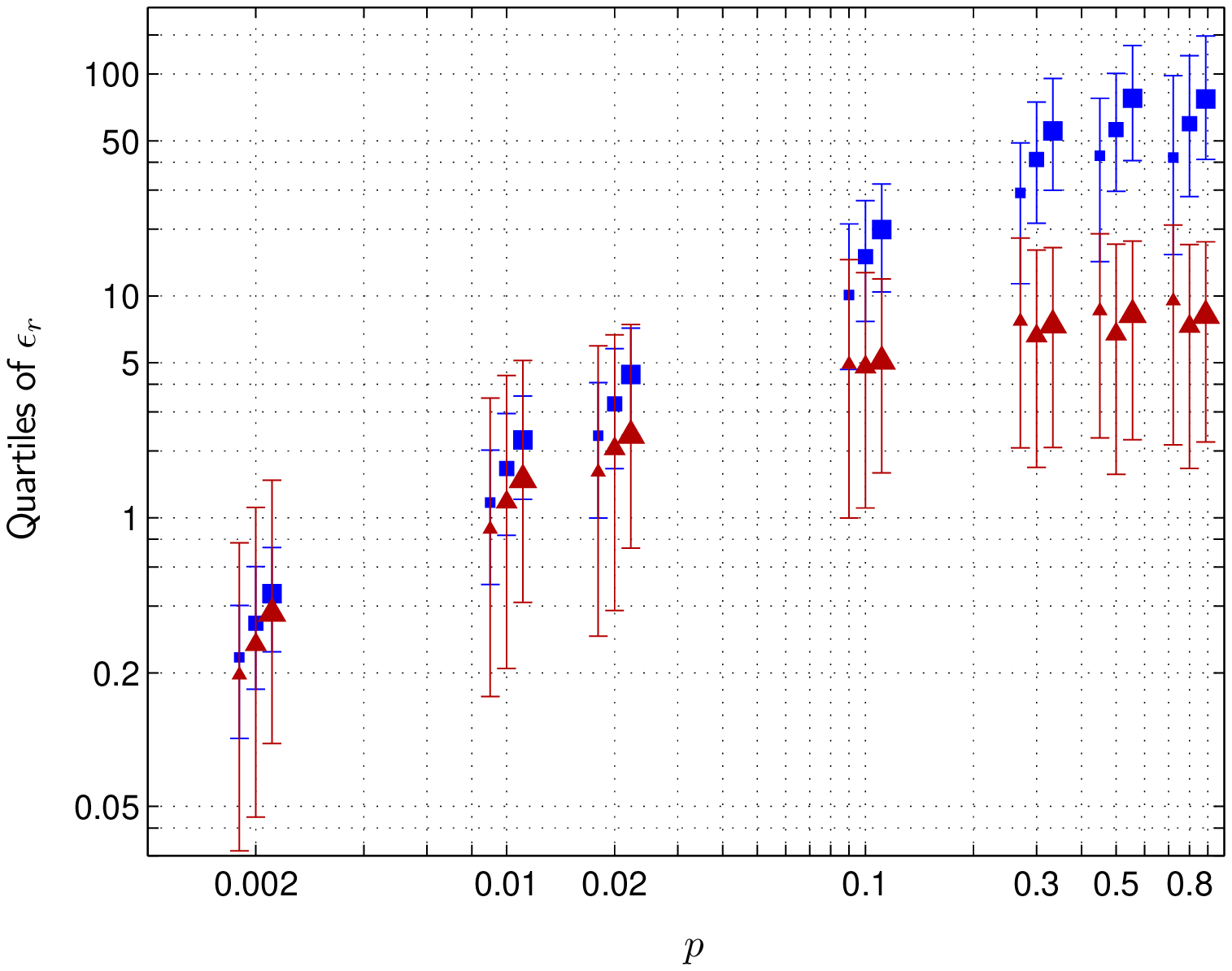} 
\caption{\textsf{Approximation of $\mu$ using (\ref{eqmue}). Quartiles of relative error $\epsilon_{r}$ (in $\%$) versus $p$ (ln-ln frame) for the six ensembles $(n,C)$. The initial condition for the calculation of $\psi_{s}$ is the uniform one. Blue squares: $C=2$; red triangles: $C=5$. The size of a symbol (square or triangle) is proportional to $n$}.}
\label{figus34}
\end{center}
\end{figure}

From (\ref{eqmoygeo}) and (\ref{eqmue2}) we get for sufficiently small $p$ that:

\begin{equation}\label{eqmue}
\mu \approx \tau / \alpha,
\end{equation} Therefore, in the low $p$ regime, the mean time spent in a basin of a \textsf{NBN} is approximately proportional to the mean specific path $d=\tau$. As shown in Fig.~\ref{figus34} for the six ensembles $(n,C)$ (ln-ln plot), the relative error $\epsilon_{r}$ done in approximation (\ref{eqmue}) decreases as $p$ becomes smaller. It can also be seen in the figure that in the low $p$ regime, $C$ being fixed, median $\epsilon_{r}$ increases with $n$, and that while at fixed $n$ the median error for chaotic basins is smaller than for complex ones, the interquartile range for the former is greater than for the latter. 

The statistical basin variables in equation (\ref{eqmue2}) are $\mu$ and $\alpha$. Since the hyperbolic function is strictly monotone decreasing, the median of $1/\alpha$ is equal to the inverse of the median of $\alpha$. Thus the constant $c_{2}$ in equation (\ref{eqhyp}) must be approximately equal to $\tilde{c}_{2}=1/ n Q_{(2;\alpha)}$, where $Q_{(2;\alpha)}$ stands for the median of $\alpha$. For $C=2$, the values of $\tilde{c}_{2}$ were found to be ($n=6$, $8$ and $10$) $0.3333$, $0.2857$ and $0.2500$; for $C=5$, we found $\tilde{c}_{2}=0.2105$, $0.1538$ and $0.1231$ respectively. These values of $\tilde{c}_{2}$ are indeed very close to the values of $c_{2}$ that have been obtained by the least squares method in subsection 4.2 (the relative error between $c_{2}$ and $\tilde{c}_{2}$ ranges from $0.25$ to $0.90\%$).

The first and third quartiles of $\mu$ satisfy a relation of the same type as (\ref{eqhyp}):

\[
Q_{(1;\mu)} \approx c_{1}/p,\quad Q_{(3;\mu)} \approx c_{3}/p,
\] where $c_{1} \approx \tilde{c}_{1} = 1/n Q_{(3;\alpha)}$ and $c_{3} \approx \tilde{c}_{3}=1/n Q_{(1;\alpha)}$. The first (resp. third) quartile of $\mu$  is thus linked to the third (resp. first) quartile of $\alpha$. For each ensemble $(n,C)$, we can define as many constants $\tilde{c}_{i}$ as there are percentiles.

\begin{figure}
\begin{center}
\includegraphics[width=1\textwidth]{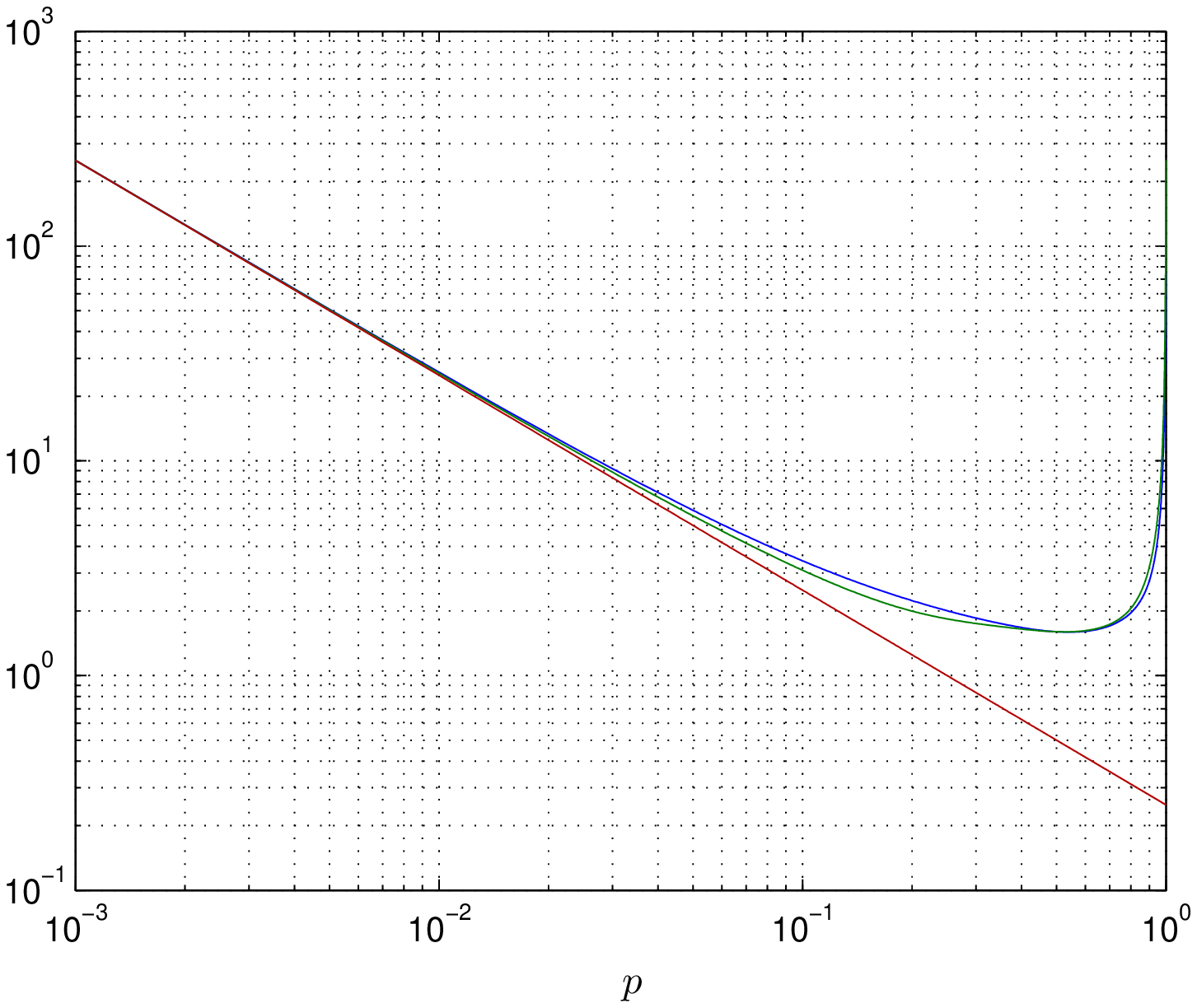}
\caption{\textsf{Comparison between $\mu$ (blue curve), $\lambda^{*}$ (green curve) and $1/n \alpha p$ (red curve) versus $p$ (ln-ln plot) for basin \textsf{B1} of Fig. \ref{figbooex} and uniform initial condition}.}
\label{figvp}
\end{center}
\end{figure}

Fig. \ref{figvp} shows $\mu$ (in blue), $\lambda^{*}$ (in green) and $1/n \alpha p$ (in red) versus $p$ (ln-ln plot) for basin \textsf{B1} of Fig. \ref{figbooex} and the uniform initial condition. For small $p$ ($p < 10^{-2}$ on the figure), these three functions behave identically. For $p=0.5$, we see that $\mu=\lambda^{*}$. For basin \textsf{B2} of Fig. \ref{figbooex} and the uniform initial condition, $\mu$ is very close to $\lambda^{*}$ whatever $0 < p < 1$ (figure not shown). Thus formula $\mu \approx \lambda^{*}$ may be accurate even if $p$ is not small, which is not the case for approximation $\mu \approx 1/n \alpha p$.

Since $p_{\mu}=1/\mu$, from (\ref{eqmoygeo}) and (\ref{eqmue2}) we can express $p_{\mu}$ through some approximation formulas valid in the low $p$ regime:

\begin{equation}\label{eqps1}
p_{\mu} \approx n p \alpha.
\end{equation} Since for sufficiently small $p$, $r \approx np$, we may write:

\begin{equation}\label{eqps3}
p_{\mu} \approx r \alpha.
\end{equation} If $\alpha$ is in the $i$th main probability level (see 4.1), then:

\begin{equation}\label{eqps2quanti}
p_{\mu} \approx i p, \quad i=1,2,\ldots,n.
\end{equation} If $\psi_{s}$ is geometric, then its parameter is equal to $p_{\mu}$. Only in this case is the probability of leaving the basin during one time step equal to the inverse of the mean sojourn time $\mu$. If $\psi_{s}$ is close to a geometric distribution (for some or all $\mathbf{b}^{(0)}$), then $p_{e}(k,k+1)$ is approximately constant, i.e. $p_{e}(k,k+1) \approx p_{\mu}=1/ \mu$. 

Let us mention another property. Let $p_{e}(k,k+A)=\sum_{j=1}^{A} p_{e}(k+j-1,k+j)$. Then

\[
\lim_{k \to \infty} p_{e}(k,k+A) = \frac{A}{\lambda^{*}}.
\] Equivalently, at fixed $p$, the mean probability of leaving a basin calculated over a period $A$ of the attractor tends to be constant as time increases. The convergence is much more rapid than the one of $p_{e}(k,k+1)$ (see Proposition 1). 



Finally, we shall establish the general expression of the mean sojourn time when $p=0.5$. The number of ways any state of an $n$-node \textsf{NBN} can be perturbed is:

\begin{equation}\label{eqbinth}
\sum_{x=1}^{n} \binom{n}{x}=2^{n}-1,
\end{equation} where the last equality follows from the Binomial Theorem. This means that whatever the perturbed state, any of the $(2^{n}-1)$ other states is reachable from that state by applying the appropriate perturbation combination out of the $(2^{n}-1)$ possible perturbation combinations (the chain is irreducible). Hence:

\[
\sum_{x=1}^{n} \Gamma_{i}^{x}=2^{n}-B \quad \forall i \in \textsf{B},
\] with, as before, $B$ the size of basin \textsf{B}. For $p=0.5$, we know that $\psi_{s}$ is geometric, i.e. the probability $p_{\mu}$ does not depend on $k$ nor on the initial conditions. Thus:

\begin{eqnarray}\label{eqps05}
p_{\mu} & = & \frac{2^{n}-B}{2^{n}} \nonumber \\
      & = & 1-\frac{B}{E},
\end{eqnarray} and thus

\begin{equation}\label{eqmu05}
\mu=\lambda^{*}=\frac{E}{E-B}.
\end{equation} The bigger the basin, the smaller $p_{\mu}$, the higher $\mu$. The Perron-Frobenius eigenvalue of $Q$ is $B/E$ and that of $(I-Q)^{-1}$ is $E/(E-B)$. When $B=1$ we get $\mu=\tau=\lambda^{*}=E/(E-1)$. Notice that, since formula (\ref{eqmu05}) depends only on $E$ and $B$, it is also valid when $\alpha=0$.

Another way to get (\ref{eqps05}) is to notice that when $p=0.5$, each element of $\Pi'$ (and $\Pi''$) is equal to $(0.5)^n=1/E$. This means that

\[
a_{i}=\frac{E-B}{E} \quad \forall i \in \textsf{B}.
\] Since $a_{i}$ does not depend on time nor on initial state $i$, $\psi_{s}$ must be geometric.

\section{Method for the reduction of a \textsf{NBN}}

\subsection{Discrete-time reduction}

\subsubsection{The low $p$ regime}

Consider an $R$-basin \textsf{NBN} with state space size $E$ and Markov representation $\mathbf{X}_{k}$ and suppose that the network has no $\alpha=0$ basin. We want to find a discrete-time homogeneous Markov chain $\{ \tilde{\mathbf{Y}}_{k}, k=0,1,2,\ldots\}$ with state $i$ of the chain representing basin $i$ of the network, i.e. the state space of $\tilde{\mathbf{Y}}_{k}$ is $\{1,2,3,\ldots,R\}$. In the theory of Markov processes, $\tilde{\mathbf{Y}}_{k}$ would be called a reduced chain or aggregated chain because $R < E$. The problem of reducing a Markov chain to a chain with a smaller state space, the so-called ``state space explosion problem'', is not new \citep{KS60,FR86} and is still an active field of research in the theory of Markov processes \citep{GU06,GR08,WE08,ZH09}. Here, we do not need to aggregate $\mathbf{X}_{k}$. The aggregation is fixed by the interactions that occur between the components of the network. 

The expressions for the transition probabilities $\tilde{\pi}_{ij}=\mathsf{Pr} \{ \tilde{\mathbf{Y}}_{k+1}=j \vert \tilde{\mathbf{Y}}_{k}=i \}$ between the basins of the network are found as follows. For convenience, the validity of these formulas are discussed further below. From (\ref{eqps1}), we write:

\[
 \tilde{\pi}_{ii} = 1 - n p \alpha_{i},
\] where $\alpha_{i}$ denotes the $\alpha$ probability of basin $i$ ($\alpha_{i} >0$, $\forall i$). Thus we have:

\begin{equation}\label{eqali2}
\sum_{j \neq i} \tilde{\pi}_{ij} = n p \alpha_{i}, \quad i=1,2,\ldots,R.
\end{equation} Probability $\alpha_{i}$ can be expressed as a sum of probabilites:

\begin{equation}\label{eqali}
\alpha_{i}=\sum_{j \neq i} \alpha_{ij},
\end{equation} with $\alpha_{ij}$ the conditional probability for a transition between basins $i$ and $j$ to occur given that one bit of attractor $i$ has been perturbed ($\alpha_{ii}=0$). If $\alpha_{ij} > 0$ ($i \neq j$), then transition probability $\tilde{\pi}_{ij}$ is taken to be:

\begin{equation}\label{eqpiprime2}
\tilde{\pi}_{ij} = n p \alpha_{ij}, \quad i \neq j.
\end{equation} Therefore in the low $p$ regime, $\tilde{\pi}_{ij}$ is the product of two probabilities: the probability that one node be perturbed during one time step, which is approximately equal to $np$ for small $p$, and the conditional probability $\alpha_{ij}$. 

Now if $\alpha_{ij}=0$, one may transition to basin $j$ by perturbing at least two bits of an attractor state or at least one bit of a transient state. Since in the low $p$ regime the network is rarely found in transient states and $L(p;n,x)=o(p)$ for $2 \leq x \leq n$, when $\alpha_{ij}=0$, transitions $i \to j$ are rare events compared to transitions $i \to j$ for which $\alpha_{ij} > 0$. Therefore we take $\tilde{\pi}_{ij}=0$.

By supposing that the time spent in any basin \textsf{B} of a \textsf{NBN} is geometric with mean $1/n p \alpha$, we neglect transitions of order $1$ from transient states as well as transitions of order $2$ or more (transitions from transient or attractor states due to perturbations affecting two or more nodes simultaneously). This means that while the original chain is irreducible, the reduced chain may not be irreducible anymore. If this is the case, the reduction method may give inaccurate results. Suppose that reduction of a \textsf{NBN} gives two sets of basins, each containing two basins that communicate with each other (each basin is accessible from the other), and that those sets are closed (by perturbing any node of any attractor state of any set, the other set cannot be reached). In Markov theory, such sets are called closed communicating classes. If we start in one set with probability one, then the state probability in the other set calculated from the reduced matrix will be $0$ at any time. Now if we solve the original chain, this will not be the case. If the stationary probability for the initially empty set is not negligible, then it will take a long time to approach this probability with good accuracy but it will. Another difference between these two chains is that the reduced one has an infinity of stationary distributions while the original one a unique stationary distribution.

If, starting in any basin, one can reach any other basin by applying single node perturbations to attractor states, then the reduced chain is irreducible. In this case, the smaller $p$, the more accurate the reduction method. If the reduced chain is not irreducible, the reduction method is not guaranteed to work properly. 

\begin{rem}
We investigated the case of reducible chains. Let $\psi_{ij}$ be the probability distribution of the time spent in basin $i$ given $\mathbf{b}_{0}$ and arrival basin $j$. When $\alpha_{ij}=0$, the first moment $\mu_{ij}$ of $\psi_{ij}$ may strongly depend on $\mathbf{b}_{0}$. We found some cases (some basins with some $\mathbf{b}_{0}$) in the low $p$ regime for which $\mu_{ij}$ was significantly smaller than $\mu$. 
\end{rem}

To illustrate the chain reduction method, we chose a randomly generated $(8,2)$ network having $R=4$ basins of size $72$, $120$, $36$ and $28$. The size of the corresponding attractors were $6$, $6$, $1$ and $3$. We considered two $p$ values, namely $0.01$ and $0.1$. The reduced matrix $\tilde{\Pi}$ when $p=0.01$ was found to be:

\begin{equation}\label{eqlambda}
\tilde{\Pi}=
\left( \begin{array}{cccc}
0.9633 & 0.0200 & 0  & 0.0167 \\
0.0133 & 0.9700 & 0.0100 & 0.0067 \\
0  & 0.0400 & 0.9600 & 0 \\
0.0333 & 0.0200 & 0 & 0.9467 \\
\end{array} \right).
\end{equation} This matrix is irreducible, i.e. any basin is accessible from any other basin by applying single-node perturbations to attractor states\footnote{Notice that there is no $\alpha=0$ basin ($\alpha_{i} > 0$ $\forall i$). If $\alpha_{i}$ was null for some $i$, we would have $0$ everywhere in row $i$ of $\tilde{\Pi}$ except at position $i$ where we would have $1$. Thus the reduced chain would be absorbing.}. Also note that the second basin, which is the biggest one ($120$ states), is the only state of $\tilde{\mathbf{Y}}_{k}$ which is reachable from any other state.

\begin{figure}
\begin{center}
\includegraphics[width=0.8\textwidth]{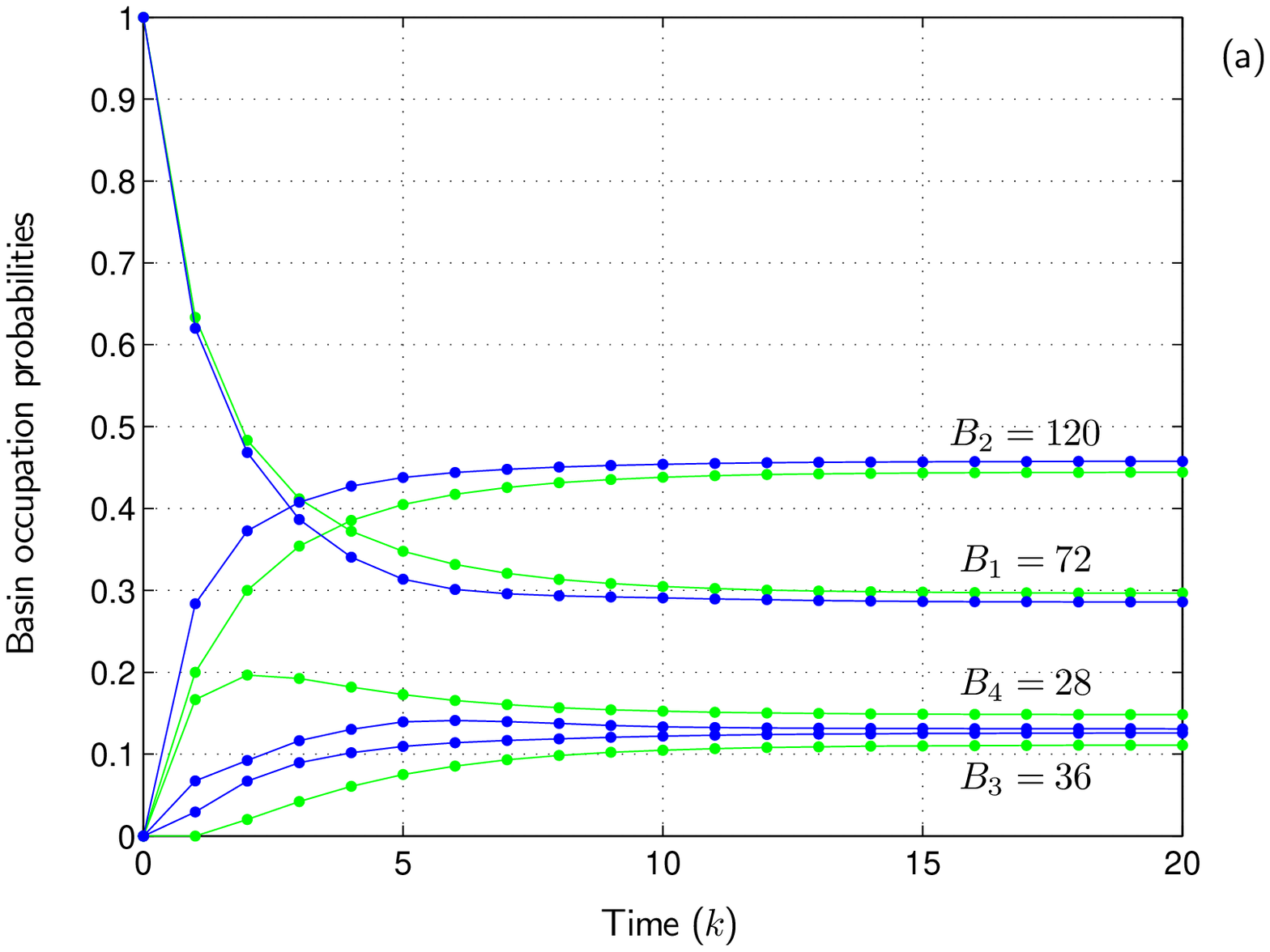} \\
\includegraphics[width=0.8\textwidth]{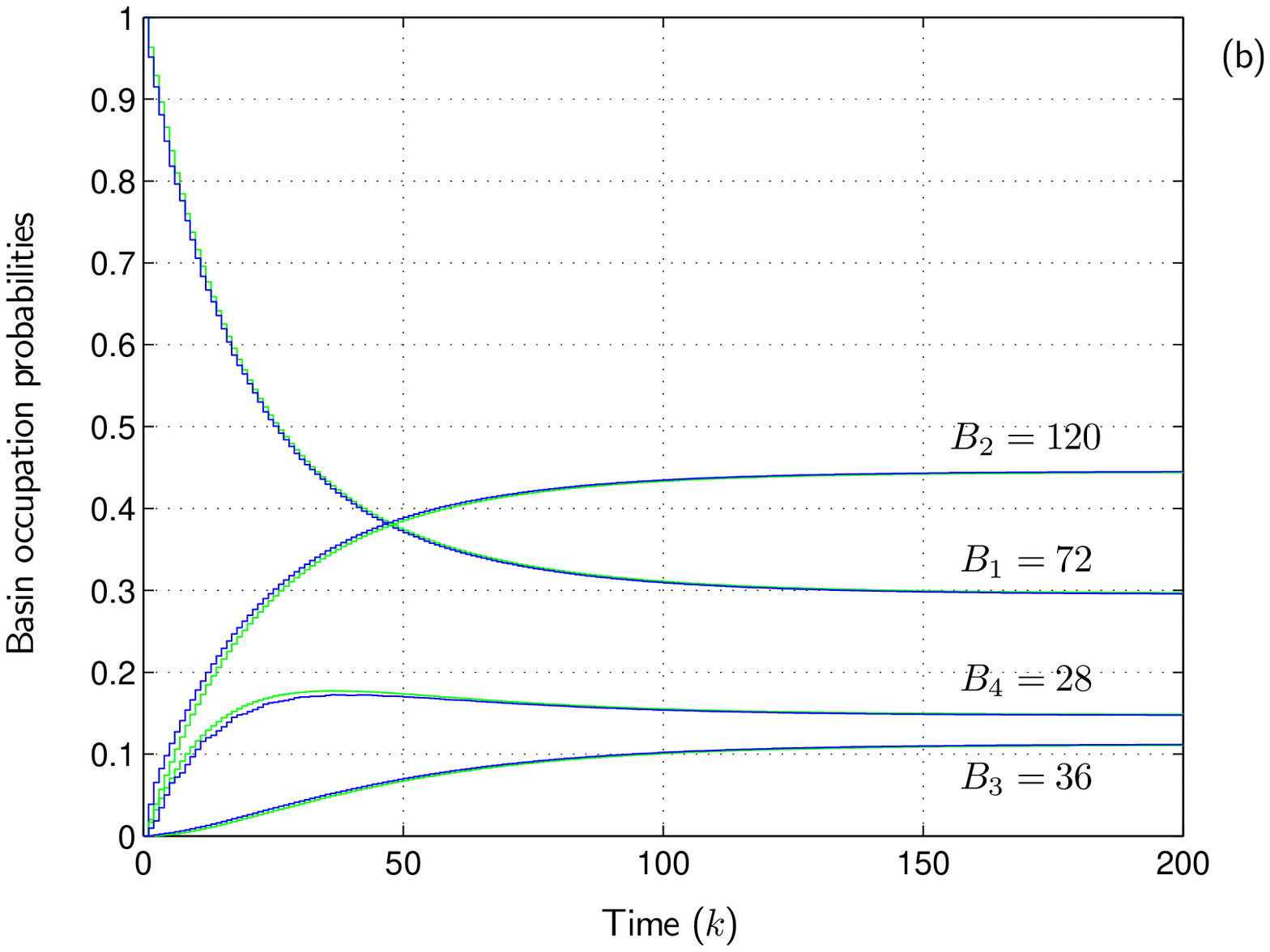} \\
\caption{\textsf{Reduction of a $(8,2)$ noisy Boolean network. In blue: basin occupation probabilities $Z_{i}^{(k)}$ calculated from matrix $\Pi$ when initially the network is in state $00000000 \in \textsf{B1}$ (state $1$ of $\mathbf{X}_{k}$); in green: state probabilities $\tilde{z}_{i}^{(k)}$ of $\tilde{\mathbf{Y}}_{k}$ calculated from matrix $\tilde{\Pi}$ (initial conditions: the chain is in state $1$ of $\tilde{\mathbf{Y}}_{k}$). Transition probabilities $\tilde{\pi}_{ij}$ are estimated from (\ref{eqpiprime2}). (a) p=0.1 (for the sake of clarity, the probabilities were interpolated linearly); (b) p=0.01 (stairstep plots and points omitted)}.}
\label{figreduc}
\end{center}
\end{figure}

The four basin occupation probabilities versus time are shown in Fig. \ref{figreduc}. The probabilities calculated from the $256 \times 256$ matrix $\Pi$ of the original chain $\mathbf{X}_{k}$, which we denote by $Z_{i}^{(k)}$, are represented in blue. At time $0$, the network was put in state $1$ (state $00000000$) which is in \textsf{B1}, so that $z_{1}^{(0)}=Z_{1}^{(0)}=1$. The state probabilities $\tilde{z}_{i}^{(k)}$ of the reduced chain $\tilde{\mathbf{Y}}_{k}$, calculated from the $4 \times 4$ matrix $\tilde{\Pi}$, are shown in green. It is seen in Fig.~\ref{figreduc}a that for $p=0.1$ approximation $\tilde{\mathbf{Y}}_{k}$ is not good, in particular for basin \textsf{B4}. Results for $p=0.01$ are presented in Fig.~\ref{figreduc}b where it can be seen that the blue and green stairstep plots are almost identical, i.e. for $p=0.01$, $\tilde{\mathbf{Y}}_{k}$ is a faithful coarse-grained representation of the \textsf{NBN}. 

Also compared the stationary probabilities calculated from $\mathbf{X}_{k}$ and $\tilde{\mathbf{Y}}_{k}$ (see 1.4). For $\mathbf{X}_{k}$, we found (basins $1$, $2$, $3$ and $4$) $0.2856$, $0.4577$, $0.1259$ and $0.1308$ when $p=0.1$;$0.2952$, $0.4455$, $0.1122$ and $0.1472$ when $p=0.01$. For $\tilde{\mathbf{Y}}_{k}$, the stationary probabilites are independent of $p$ (see below) and equal to $0.2963$, $0.4444$ $0.1111$ and $0.1481$. The maximum of the relative error is $13.26\%$ when $p=0.1$ and $0.94\%$ when $p=0.01$. Thus in the long run, the most populated basin is the one with the greatest size ($120$) and the smaller $\alpha$ probability ($3/8$).

More generally, if chain $\tilde{\mathbf{Y}}_{k}$ is irreducible and aperiodic then its stationary state probability vector satisfies:

\begin{equation}\label{eqpired}
\tilde{\mathbf{z}} = \tilde{\mathbf{z}} \tilde{\Pi} \qquad \sum_{i} \tilde{z}_{i}=1.
\end{equation} Rearranging the first equation in (\ref{eqpired}) we find:

\[
A_{\alpha} \tilde{\mathbf{z}}=0, \quad \tilde{z}_{1}+\tilde{z}_{2}+\ldots+\tilde{z}_{R}=1,
\] where

\begin{equation}
A_{\alpha} =
\left( \begin{array}{cccccccc}
 -\alpha_{1} & \alpha_{21} & \alpha_{31}  & \ldots  & \alpha_{R1}  \\
 \alpha_{12} & -\alpha_{2} & \alpha_{32}  & \ldots  & \alpha_{R2}  \\
 \alpha_{13} & \alpha_{23} & -\alpha_{3}  & \ldots  & \alpha_{R3}  \\
 \vdots      &    \vdots   &       \vdots & \ddots  &  \vdots      \\
\alpha_{1R} & \alpha_{2R} & \alpha_{3R} &     \ldots   & -\alpha_{R}
\end{array} \right).
\end{equation} From (\ref{eqali}), matrix $A_{\alpha}$ is singular. The stationary state probability vector of $\tilde{\mathbf{Y}}_{k}$ is thus an eigenvector of $A_{\alpha}$ and the corresponding eigenvalue is $0$. 
 
In addition, the stationary state probability vector can be obtained by calculating the limit

\begin{equation}\label{eqpistatred}
\lim_{k \to \infty} (I+A_{\alpha})^{k},
\end{equation} where each row of the limit matrix is equal to the transposed stationary state probability vector.  

\subsubsection[Case $p=1/2$]{Case $p=0.5$}

When $p=0.5$, the reduction is ``exact''. From (\ref{eqps05}), the probability to leave a basin of size $B$ is $1-B/E$. The probability to remain in such a basin is thus $B/E$. Since $\psi_{s}$ is geometric, one has:

\begin{equation}
\tilde{\pi}_{ij} = (1-\frac{B_{i}}{E}) w_{ij} \quad \textrm{if} \quad i \neq j,
\end{equation} with

\[
w_{ij}=\frac{B_{j}}{\displaystyle \sum_{j \neq i} B_{j}},
\] and $B_{i}/{E}$ otherwise. Hence:

\begin{equation}
\tilde{\pi}_{ij} = \frac{B_{j}}{E}, \quad i=1,2,\ldots,R.
\end{equation} The rows of $\tilde{\Pi}$ are thus equal. In fact, when $p=0.5$, the stationary state is reached in one step whatever the initial conditions so that each row of $\tilde{\Pi}$ gives the stationary state probabilities for the basins of the network. Note that since $\delta^{*} \to 0$ as $p \to 0.5$, these stationary state probabilities can be used to approximate the stationary state probabilities in a neighborhood of $p=0.5$. 

From $\tilde{\Pi}$, the mean sojourn time in basin $i$ is expressed as:

\[
\mu_{i}=\frac{1}{1 - \tilde{\pi}_{ii}}=\frac{E}{E-B_{i}}.
\] 

\subsection{Continuous-time reduction}

The preceding sections deal with discrete-time homogeneous Markov chains. For such chains, the sojourn time spent in any state is geometric and thus memoryless. The continuous analog of the geometric distribution is the exponential distribution, which is also memoryless. Making the passage from geometric to exponential distribution leads to continuous-time homogeneous Markov chains. As we shall discuss, in the continuous-time representation of Markov processes, the state probabilities satisfy a system of linear ordinary differential equations.

A continuous-time Markov chain $\{ \tilde{\mathbf{Y}}_{t}, t \geq 0 \}$ is homogeneous if the transition probability from state $i$ to state $j$ in time interval $(t,t+\Delta t)$ depends only on the length $\Delta t$ of the interval: $\tilde{\pi}_{ij}(t,t+\Delta t) = \mathsf{Pr} \{ \tilde{\mathbf{Y}}_{t+\Delta t}=j \vert \tilde{\mathbf{Y}}_{t}=i \} = \tilde{\pi}_{ij}(\Delta t)$. In this case \citep{KL75}:

\begin{equation*}
\tilde{\pi}_{ij}(\Delta t) =  u_{ij} \Delta t + o(\Delta t) \quad \textrm{if} \quad i \neq j \quad \textrm{and}
\end{equation*}
\begin{equation}\label{eqcont1}
1 - \tilde{\pi}_{ii}(\Delta t) =  u_{ii} \Delta t + o(\Delta t),
\end{equation} with $u_{ij} \geq 0$ the rate of transition from state $i$ to state $j \neq i$ and 

\begin{equation}\label{eqcont3}
\sum_{j \neq i} u_{ij} = u_{ii}.
\end{equation} The time spent in state $i$ is exponentially distributed with parameter $u_{ii}$.



Now if $\tilde{\mathbf{Y}}_{t}$ is viewed as the continuous-time coarse-grained representation of an $R$-basin \textsf{NBN}, then the state probabilities $\tilde{z}_{i}(t)$ of $\tilde{\mathbf{Y}}_{t}$ satisfy the following Master equation \citep{KL75}:

\begin{equation}\label{eqcontinu}
\frac{d}{dt} \tilde{z}_{i}(t) = -u_{ii} \tilde{z}_{i}(t) + \sum_{j \neq i} u_{ji} \tilde{z}_{j}(t), \quad i=1,2,\ldots,R.
\end{equation} As stated in Proposition~3, to allow the passage from discrete to continuous representation, only one needs $p$ to be sufficiently small. In other words, for sufficiently small $p$, one may use $\tilde{\mathbf{Y}}_{t}$ instead of $\tilde{\mathbf{Y}}_{k}$. A small $p$ implies that the mean specific path $d=\tau$ is large compared to $1$. Since $\mu \geq \tau$, a small $p$ also implies that $\mu$ is large compared to $1$. As an example, compare Figs. \ref{figreduc}a and \ref{figreduc}b. In the first case, $d=1.7558$ and $\mu \approx 3$ whatever the basin, while in the second case, $d=12.9441$ and $\mu$ is between $20$ and $30$ depending on the basin.

\begin{exmp}
Let us illustrate the passage from discrete-time chain $\mathbf{X}_{k}$ to continuous-time reduced chain $\tilde{\mathbf{Y}}_{t}$ with the state diagram of Fig. \ref{figbooex}. The reduced chain in this example has only two states so that from (\ref{eqcont3}) we get: $u_{11}=u_{12}=u_{1}$ and $u_{22}=u_{21}=u_{2}$. The equations for the reduced chain are thus:

\begin{eqnarray}\label{eqcont4}
\frac{d\tilde{z}_{1}}{dt} & = & -u_{1} \tilde{z}_{1} + u_{2} \tilde{z}_{2} \nonumber \\
\frac{d\tilde{z}_{2}}{dt} & = &  u_{1} \tilde{z}_{1} - u_{2} \tilde{z}_{2}.
\end{eqnarray} The expressions for the transition rates are found from (\ref{eqpiprime2}) and (\ref{eqcont1}): taking $\Delta t=1$, we get $u_{1} = \tilde{\pi}_{12} / \Delta t = n p \alpha_{12} = np$ and $u_{2}= \tilde{\pi}_{21} / \Delta t = n p \alpha_{21} = np/2$. 

Figs.~\ref{figmardiff}a and \ref{figmardiff}b show \textsf{B1} occupation probability versus time when $p=0.02$ and $p=0.002$ respectively. The blue points represent the solutions of equation (\ref{eqmarkov}) when initially the states of \textsf{B1} are equiprobable and \textsf{B2} is empty (for the sake of clarity, the points were interpolated linearly) while the green continuous curves are the solutions of system (\ref{eqcont4}) when $\tilde{z}_{1}(0)=1$ and $\tilde{z}_{2}(0)=0$. For $p=0.002$ ($\mu=125.3756$), the probability calculated from $\Pi$ decreases by small amounts and seems to vary continuously with time (see the enlarged portion in Fig. \ref{figmardiff}b) so that continuous-time chain $\tilde{\mathbf{Y}}_{t}$ may be used instead of $\tilde{\mathbf{Y}}_{k}$. Thus for sufficiently small $p$, system of differential equations (\ref{eqcont4}) may be used as a coarse-grained representation of the \textsf{NBN}.  

To end this example, let us estimate the relative error between the inverse of the mean sojourn time $p_{\mu}=1/ \mu$ and its approximation $n \alpha p$. The mean sojourn time for both basins and both $p$ values are given in Table \ref{tabdis}. For $p=0.02$ one finds $1/\mu_{1}= 0.0749$ and $1/\mu_{2}=0.0369$, to be compared to $np=0.08$ and $np/2=0.04$, which gives relative errors of $6.81$ and $8.40\%$. For $p=0.002$ one gets $1/\mu_{1}= 0.0079$ and $1/\mu_{2}=0.0040$, to be compared to $np=0.008$ and $np/2=0.004$, which gives relative errors of $0.66$ and $0\%$.

\begin{figure}
\begin{center}
\includegraphics[width=0.8\textwidth]{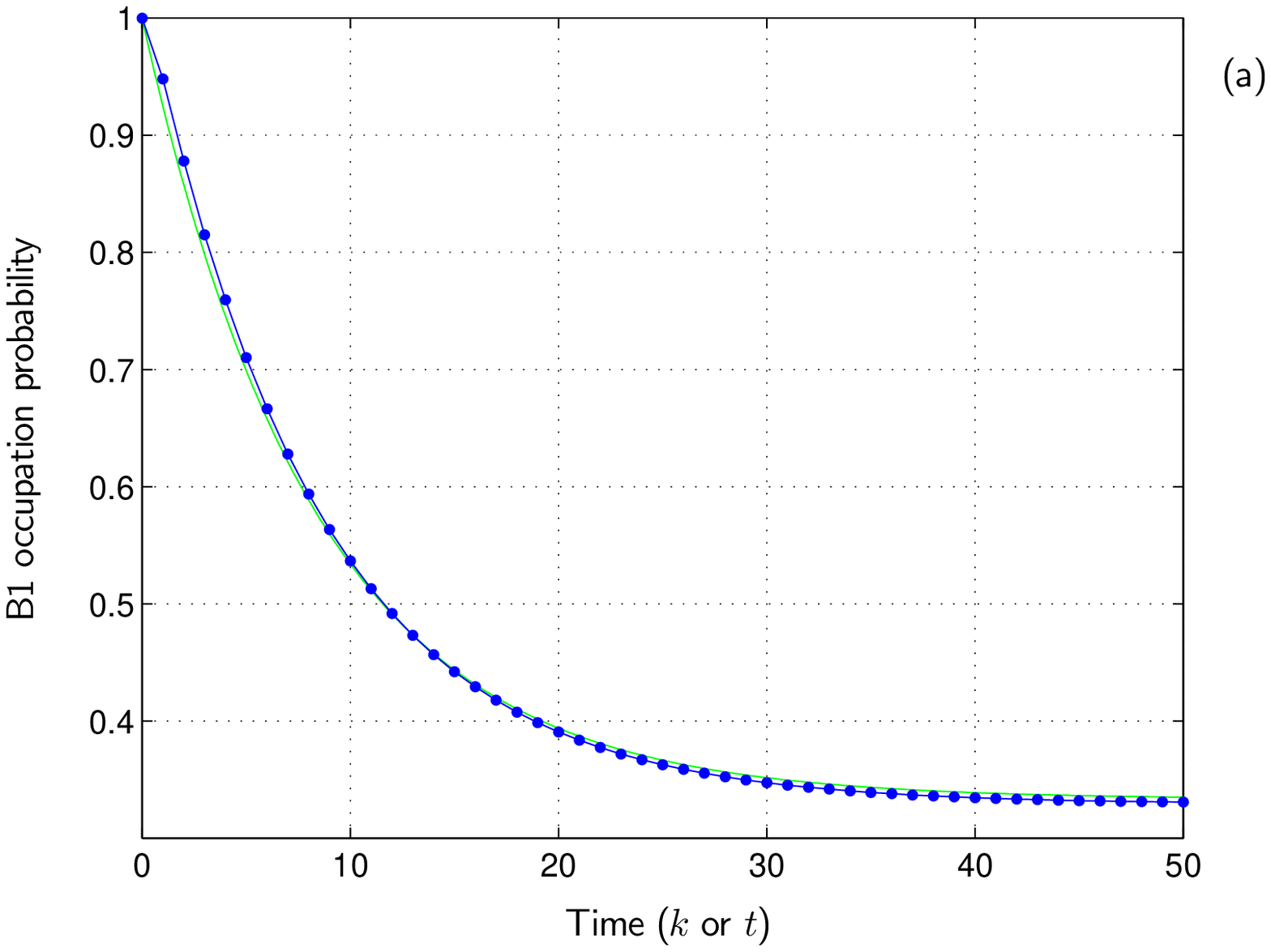} \\
\includegraphics[width=0.8\textwidth]{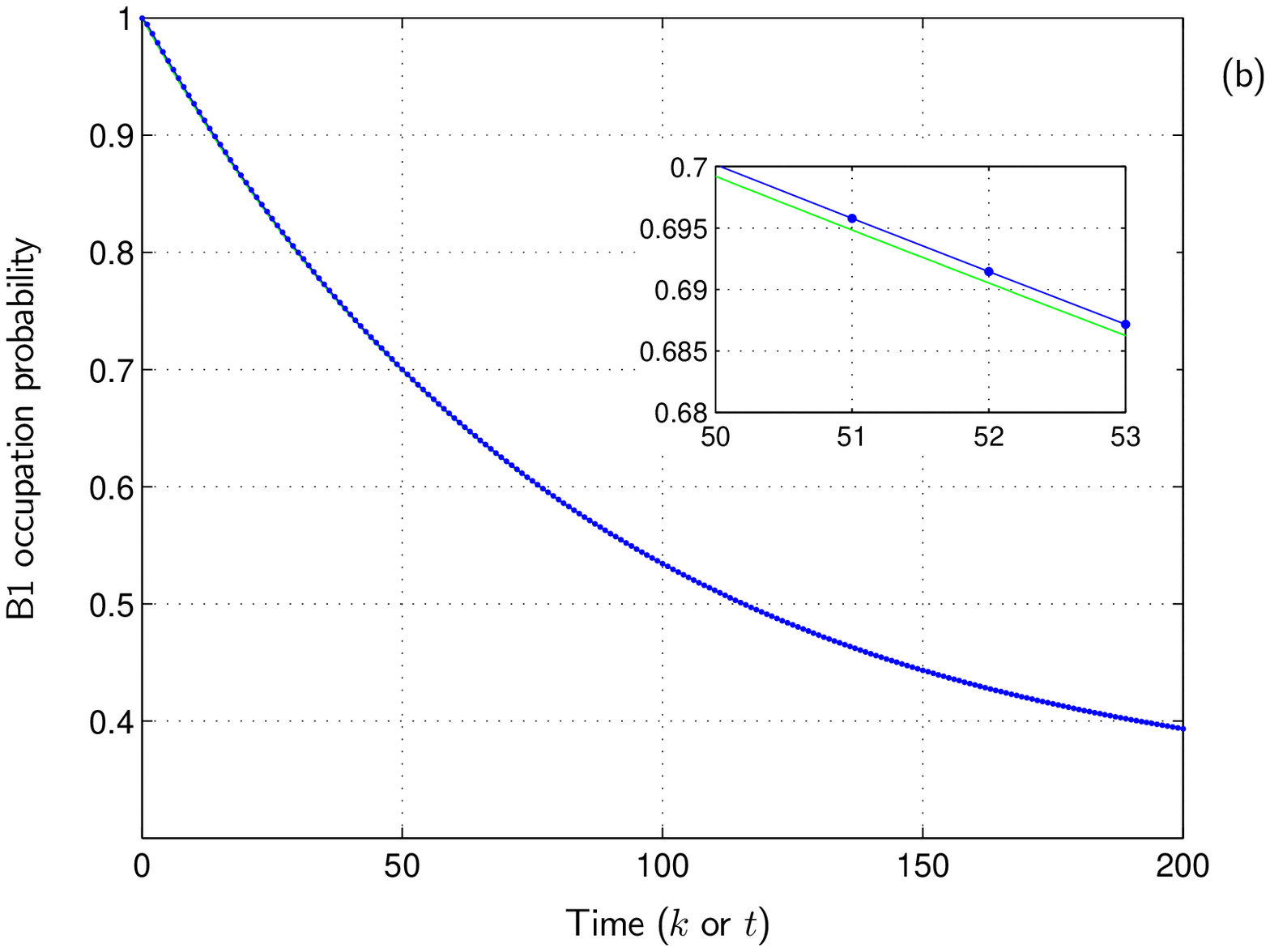} \\
\caption{\textsf{Passage from discrete-time chain $\mathbf{X}_{k}$ to continuous-time reduced chain $\tilde{\mathbf{Y}}_{t}$. Illustration with the state diagram of Fig.~\ref{figbooex}. Ordinate: B1 occupation probability; abscissa: time. Blue points: solution of (\ref{eqmarkov}) when initially the states of B1 are equiprobable (the states of B2 are empty). For the sake of clarity, the relative frequencies were interpolated linearly; solid green line: solution of equations (\ref{eqcont4}) with $u_{1}=np \alpha_{12}=np$, $u_{2}=np \alpha_{21}=np/2$, $\tilde{z}_{1}(0)=1$ and $\tilde{z}_{2}(0)=0$. (a): p=0.02; (b): p=0.002}.}
\label{figmardiff}
\end{center}
\end{figure}
\end{exmp}

\section{Statistical fluctuations in \textsf{NBN}s}

Consider $N$ replicas of a given \textsf{BN}, i.e. $N$ cells expressing the same biochemical network, and suppose that each node of each replica may be perturbed with probability $0<p<1$ independently of time, of the other nodes and of the other replicas\footnote{Statistical independence between the replicas means that whether at time $(k+1)$ a replica has been perturbed or not does not depend on whether between $0$ and $k$ other replicas have been perturbed or not.}. 

The Markov chain model $\mathbf{X}_{k}$ is a probabilistic model. Knowing the current state of a replica, this model allows to compute the probability of finding the replica in any given state of the network at any subsequent time. Therefore, even if the initial state of the replica is known with certainty, its trajectory in the state space of the network cannot be predicted with certainty. The same applies to the prediction of the number of replicas in each state of the network at any time.

\begin{figure}
\begin{center}
\includegraphics[width=0.8\textwidth]{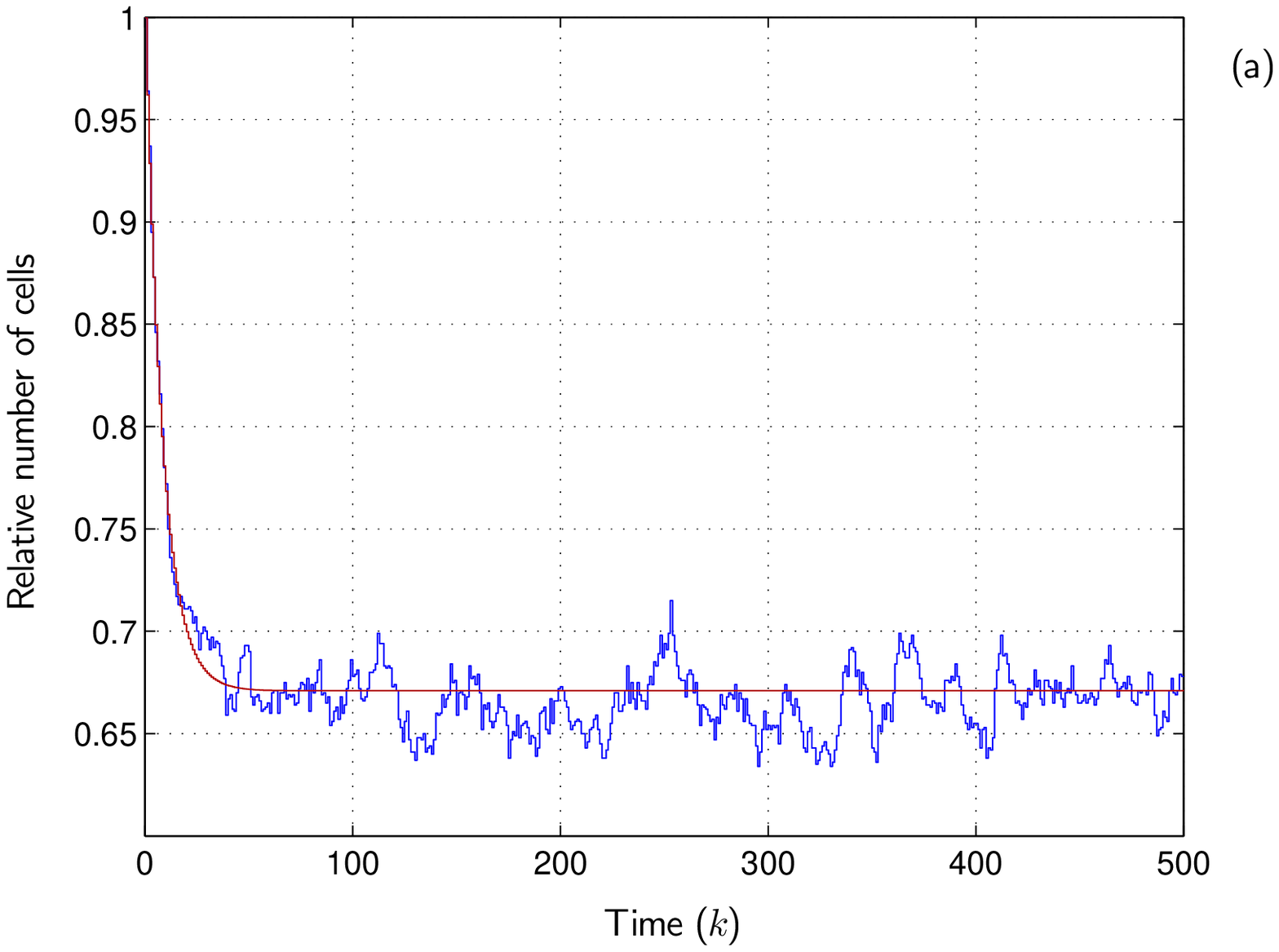} \\
\includegraphics[width=0.8\textwidth]{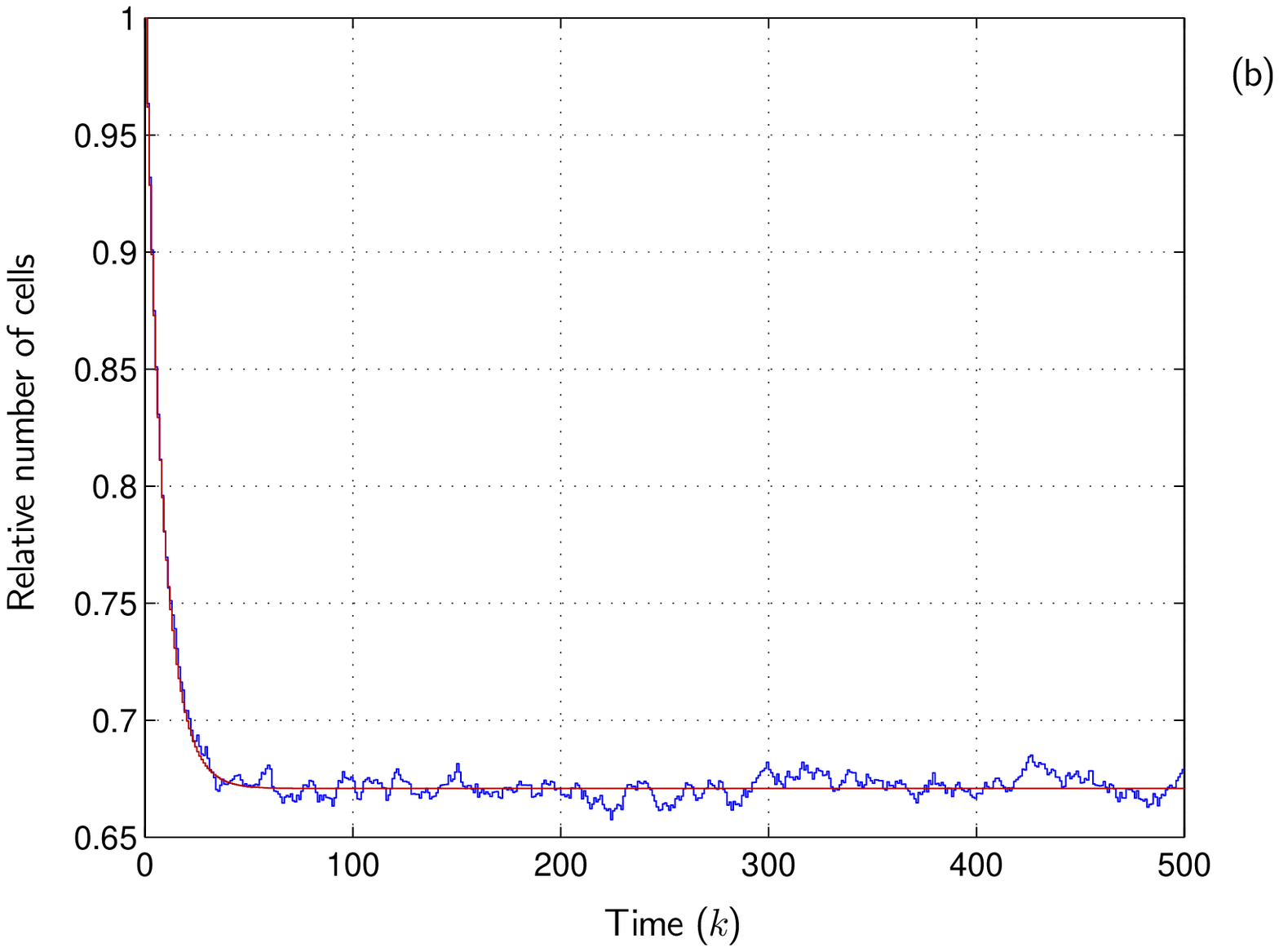} \\
\caption{\textsf{Statistical fluctuations in noisy Boolean networks. One considers $N$ replicas of the Boolean network shown in Fig.~\ref{figbooex} ($N$ cells expressing the same biochemical network). Initially, the $N$ replicas are in state $0010 \in \textsf{B2}$. The graphs show the relative number of replicas versus time when $p=0.02$ and (a) $N=10^{3}$ or (b) $N=10^{4}$. Blue stairstep plot: Monte Carlo method. Red stairstep plot: solution of matrix equation (\ref{eqmarkov})}.}
\label{figfluc1}
\end{center}
\end{figure}

In order to illustrate the random behaviour of an ensemble of replicas, random trajectories were simulated in the state space of the network of Fig. \ref{figbooex} by the Monte Carlo method. At time $0$, $N$ replicas were put in state $0010 \in \textsf{A2}$ then the number of replicas in each basin of the network at discrete times $k=1,2,\ldots$ computed. The relative number of replicas in \textsf{B2} versus time is shown in Fig.~\ref{figfluc1} for $p=0.02$ and two values of $N$. Graph (a) corresponds to $N=10^{3}$ while graph (b) to $N=10^{4}$. Each blue stairstep plot results from $N$ Monte Carlo simulations (one simulation is one trajectory of one replica), while each red one represents the mean solution calculated from (\ref{eqmarkov}). As can be seen from the two graphs, the uncertainty on the long-term behaviour of the ensemble is quite low in both cases (coefficient of variation: $\approx 2\%$ when $N=10^{3}$ and $\approx 0.7\%$ when $N=10^{4}$). 
\begin{rem}
It is assumed that the total number of cells is conserved (cells do not proliferate and they are not lost): $N_{1}^{(k)}+N_{2}^{(k)}=N \quad \forall k=0,1,\ldots$
\end{rem}

\begin{figure}
\begin{center}
\includegraphics[width=0.8\textwidth]{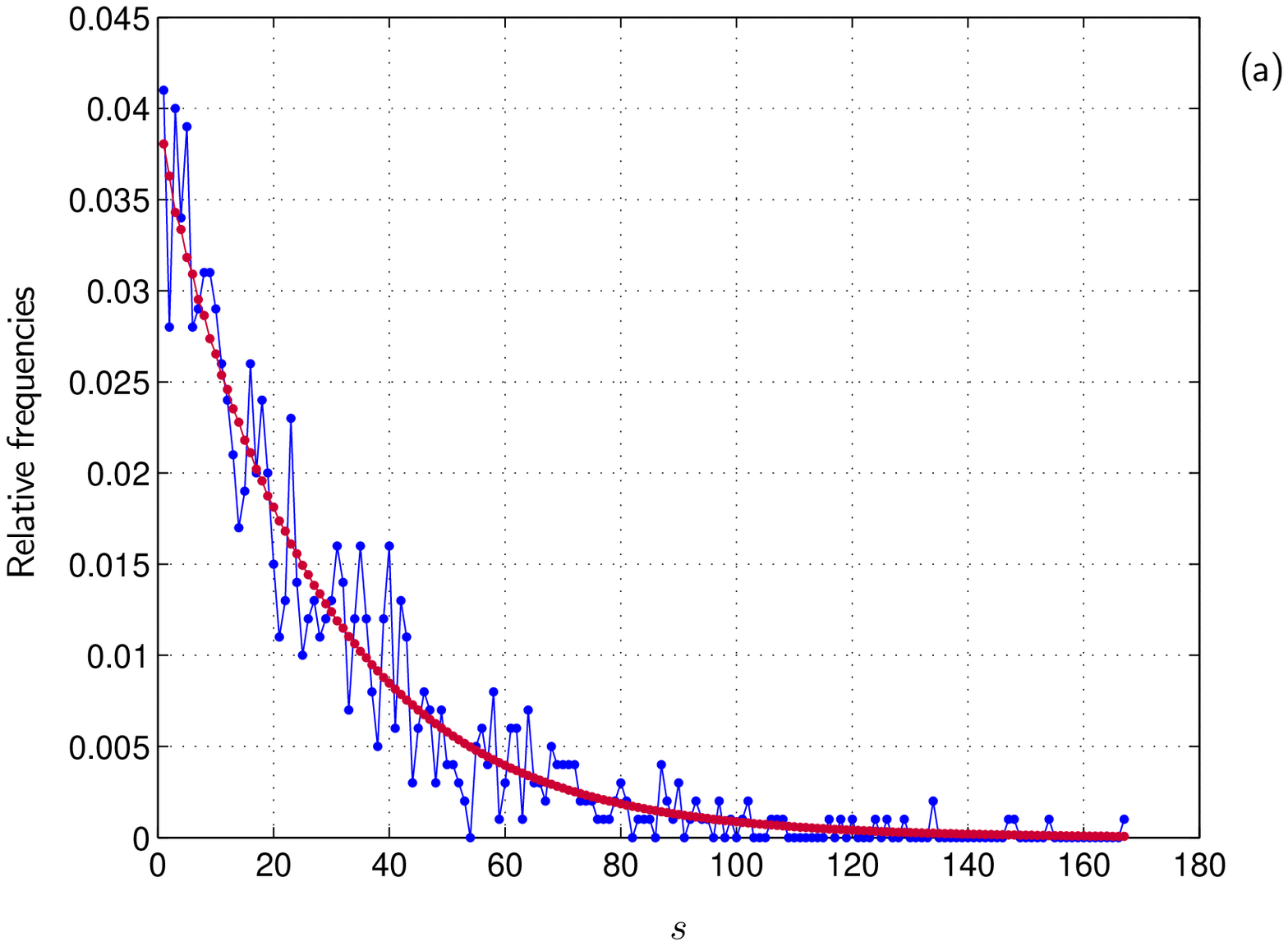} \\
\includegraphics[width=0.8\textwidth]{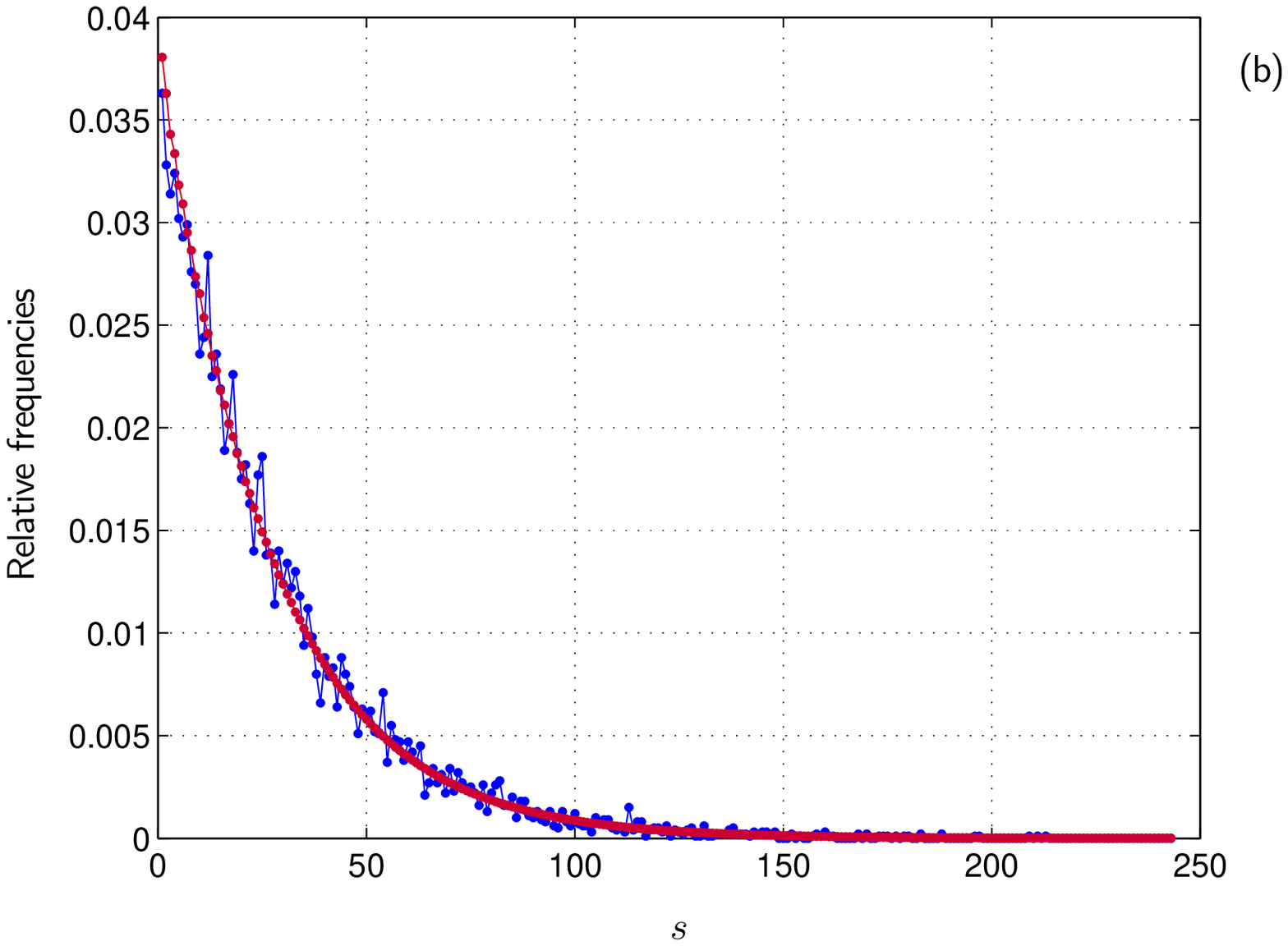} \\
\caption{\textsf{Statistical fluctuations in noisy Boolean networks. Probability distribution of the time spent in basin B2 of Fig.~\ref{figbooex} when $p=0.02$ and initially the $N$ replicas are in state $3$ (state $0010$). Blue points: Monte Carlo method with (a) $N=10^{3}$ or (b) $N=10^{4}$. Red points: Markov method, exact frequencies ($\mu=26.81$, $\delta^{*}=0.11\%$). For the sake of clarity, the relative frequencies were interpolated linearly}.}
\label{figfluc2}
\end{center}
\end{figure}

Also calculated was the probability distribution of the time spent in basin \textsf{B2}. The relative frequencies obtained from the Monte Carlo method are shown in blue in Fig. \ref{figfluc2} for $N=10^{3}$ and $N=10^{4}$ cases. The red points represent the exact frequencies calculated from matrix $Q$ as explained in section 2 (Markov method). For the sake of clarity, frequencies were interpolated linearly. Suppose that each Monte Carlo distribution in Fig. \ref{figfluc2} results from the measurement of $N$ individual sojourn times. What would be the uncertainty on the mean time spent in \textsf{B2} for each ensemble of cells ? The $95\%$ confidence interval would be $27.2760 \pm 5.8877\%$ with the $N=10^{3}$ case and $26.9015 \pm 1.9153\%$ with the $N=10^{4}$ one. The exact mean sojourn time is $26.8067$ (Markov method, see Table \ref{tabdis}).

\begin{figure}
\begin{center}
\includegraphics[width=0.8\textwidth]{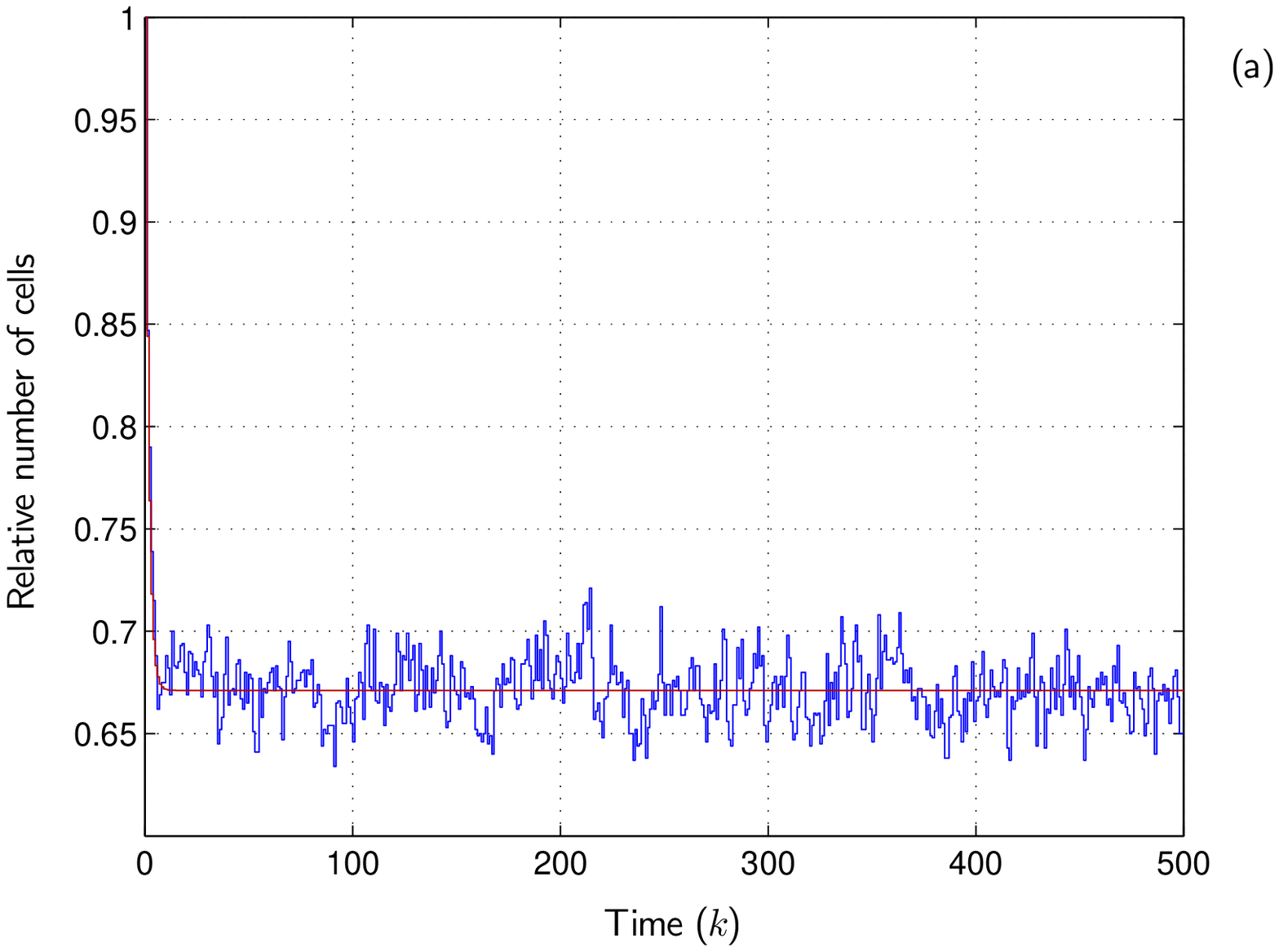} \\
\includegraphics[width=0.8\textwidth]{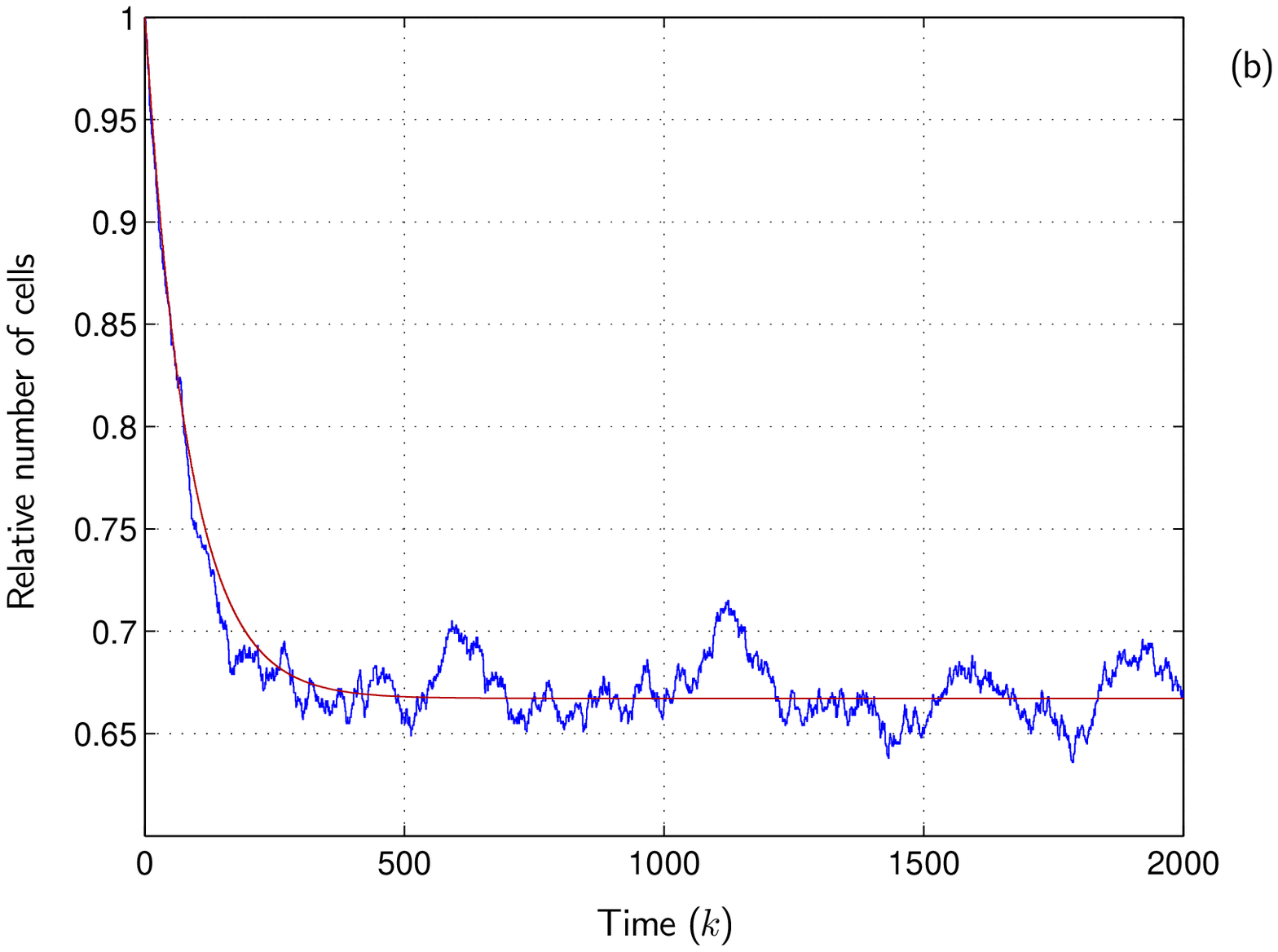} \\
\caption{\textsf{Statistical fluctuations in noisy Boolean networks. Idem Fig.~\ref{figfluc1}a except that (a) $p=0.1$ and (b) $p=0.002$. In the low $p$ regime, cells are trapped by attractors for a substantial time. Each attractor is alternatively overpopulated and underpopulated with regard to the stationary mean cell number}.}
\label{figfluc3}
\end{center}
\end{figure}

When $N$ is fixed and $p$ increases, the amplitude of the fluctuations in $\psi_{s}$ decreases as more and more replicas have a short sojourn time. By comparison of the graph of Fig. \ref{figfluc1}a and the two graphs of Fig. \ref{figfluc3}, the amplitude of the fluctuations in the relative number of replicas is not very sensitive to $p$\footnote{Neither is the long-term \textsf{B2} occupation probability: $0.6672$, $0.6709$ and $0.6711$ when $p=0.002$, $0.02$ and $0.1$ respectively.}. There is, however, a striking difference between Fig. \ref{figfluc3}a and Fig. \ref{figfluc3}b. When $p$ is small, the relative number of replicas in \textsf{B2} remains above or below its stationary value for long periods, i.e. the width of the fluctuations increases when $p$ decreases. This means that in the low $p$ regime, the long-term behaviour of the network is characterized by slow transitions between two states: one that corresponds to an ``overpopulated'' attractor and the other to an ``underpopulated'' one. The probability of crossing the stationary value during one time step was found to be $0.3111$ when $p=0.1$ and $0.0494$ when $p=0.002$ ($0.1613$ when $p=0.02$). The maximum number of time steps the attractor remains overpopulated was $16$ when $p=0.1$ and $318$ when $p=0.002$ ($54$ when $p=0.02$). Similar values were found for the underpopulated case.

\begin{rem}
(1) Since the number of replicas is conserved, when an attractor is overpopulated, the other is underpopulated and vice versa. (2) These slow transitions occuring in the low $p$ regime cannot be deduced from the Markov chain model. 
\end{rem}

\section{Reduction of a \textsf{NBN} to a two-state Markov chain}

We addressed the problem of reducing a chain $\mathbf{X}_{k}$ to a two-state homogeneous chain by aggregating basins of attraction. 

Only $(8,2)$ networks with $R \geq 4$ were studied. The aggregation process consisted in the following. When $R$ was pair, $R/2$ basins were picked at random and aggregated, while when $R$ was odd $(R-1)/2$ basins were aggregated randomly. In both cases the remaining basins were aggregated, constituting the second state of the two-state chain. For each aggregated state then, we calculated the mean sojourn time $\mu$ with both types of initial conditions (the uniform and the random type) as well as the maximum deviation $\delta^{*}$. Results indicate that as $p \to 0$, $\mu$ does not tend to be independent of initial conditions neither does $\delta^{*}$ tend to $0$. Notice, however, that the reduction of $\mathbf{X}_{k}$ worked well in some cases (some networks with some basin aggregations).


\section{Conclusion}

The reduction method for \textsf{NBN}s presented in this paper raises the important question whether biochemical networks can be reduced to (approximating) coarse-grained networks functionally equivalent to the original ones. Reducing the complexity of biochemical networks could help in the analysis of cell responses to inputs (or cell fates) by neglecting molecular interactions while focusing on the higher-level processes that emerge from those interactions. A formally equivalent and very useful reduction theorem exists in electrical circuit theory which is Thevenin's theorem. 

\section{Note}

This work is part of a manuscript entitled ``\emph{Mathematical modeling of cellular processes: from molecular networks to epithelial structures}'' written by F. Fourré. The complete manuscript contains five chapters. The first chapter is devoted to \textsf{NBN}s. The aim of the project is to propose a physical framework for describing cellular processes. Since 1st December 2008, F. Fourré has been working on a PhD thesis that is funded by the University of Luxembourg and supervised by Prof. Thomas Sauter. The thesis deals with qualitative modeling of signaling networks. 

D. Baurain is a Postdoctoral Researcher of the FNRS.

\section{Acknowledgment}

F. Fourré gratefully thanks Prof. T. Sauter for having encourage him to write this paper and for the position of Assistant/PhD student at the Systems Biology Group of the University of Luxembourg.

\bibliographystyle{apalike}
\bibliography{nbnfourre}


\end{document}